\documentclass{article}

\usepackage{arxiv}

\usepackage[utf8]{inputenc} % allow utf-8 input
\usepackage[T1]{fontenc}    % use 8-bit T1 fonts
\usepackage{hyperref}       % hyperlinks
\usepackage{url}            % simple URL typesetting
\usepackage{booktabs}       % professional-quality tables
\usepackage{amsfonts}       % blackboard math symbols
\usepackage{amsmath}
\usepackage{amssymb}
\usepackage{bbold}          % for indicator 1
\usepackage{nicefrac}       % compact symbols for 1/2, etc.
\usepackage{microtype}      % microtypography
\usepackage{graphicx}
\usepackage{natbib}
\usepackage{doi}
\usepackage[usenames,dvipsnames]{color}
\usepackage{xcolor}

\usepackage{listings}

\colorlet{lightorange}{orange!05}
\title{Extending the \emph{saemix} package for \emph{R} to fit non Gaussian outcomes}

\date{February 27th, 2026}	% Here you can change the date presented in the paper title
\author{ \href{https://orcid.org/0000-0002-9150-9886}{\includegraphics[scale=0.06]{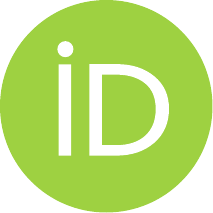}\hspace{1mm}Emmanuelle Comets}\thanks{Corresponding author} \\
Université Paris Cité \& \\
Université Sorbonne Paris Nord\\ 
Inserm, IAME, F-75018 Paris, France\\
Univ Rennes, Inserm, EHESP \\
Irset UMR\_S 1085 \\
F-35000 Rennes, France \\
	\texttt{emmanuelle.comets@inserm.fr} \\
	\And
	\href{https://orcid.org/0000-0002-3775-9278}{\includegraphics[scale=0.06]{orcid.pdf}\hspace{1mm}Maud Delattre} \\
	 Université Paris-Saclay \\
     INRAE, MaIAGE \\
     78350 Jouy-en-Josas, France \\ 
	\texttt{maud.delattre@inrae.fr} \\
	\And
	\href{https://orcid.org/0000-0000-0000-0000}{\includegraphics[scale=0.06]{orcid.pdf}\hspace{1mm}Belhal Karimi} \\
	Ecole Polytechnique \\
	Route de Saclay, Palaiseau \\
    France \\
	\texttt{belhal.karimi@polytechnique.edu} \\
}

\renewcommand{\headeright}{saemix package for non-Gaussian outcomes}
\renewcommand{\shorttitle}{{\sf arXiv} preprint}

\hypersetup{
pdftitle={saemix3},
pdfauthor={Emmanuelle Comets, Maud Delattre, Belhal Karimi},
pdfkeywords={Non-linear mixed effect models, SAEM algorithm, saemix, R package, parameter estimation},
}

\newcommand{\Eco}{\textcolor{blue}{\bf Eco: }\textcolor{blue}}

\newcommand{\Lucie}{\textcolor{violet}{\bf Lucie: }\textcolor{violet}}

\begin{document}
\maketitle

\paragraph{Corresponding author:} Emmanuelle Comets \\
\begin{tabular}{p{3.2cm} l}
&INSERM - UMR1137 \\
&UFR de Médecine Site Bichat \\
&16 rue Henri Huchard \\
& 75018 Paris, France \\
&Tel: (33) 6 25 82 49 50\\
&\texttt{emmanuelle.comets@inserm.fr} \\
\end{tabular}

\newpage
\begin{abstract} % ToDo: 350 words
{\bf Background and Objectives:} Longitudinal data are increasingly collected in clinical trials to enhance the information on treatment action and disease evolution. The trajectory of continuous biomarkers such as target hormone concentrations or viral loads can then be modelled in relationship to the occurrence of events such as recovery or hospitalisation. However many other types of longitudinal data can be collected, such as pain scores (categorical or binary), number of episodes (count) or occurrence of events (survival). The statistical analysis of choice for longitudinal data is non-linear mixed-effect models (NLMEM), as these can naturally handle individual differences in trajectories while modelling the underlying population evolution. While there are several commercial packages available to estimate the parameters of these models, such as Nonmem and Monolix, the {\sf saemix} package for {\sf R} is one of the few open-source solutions and the most flexible. In this paper, we present an extended version of the package that accommodates a variety of models for non-Gaussian data.
%In this paper, we present various models for non-Gaussian data.

{\bf Methods:} The {\sf saemix} package estimates parameters through the Stochastic Approximation Expectation-Maximisation (SAEM) algorithm. Within the package, non-Gaussian models are specified by their log-likelihood functions, which afford the user maximal control over model formulation.
%For non-Gaussian models, models are defined via their log-likelihood, which gives the user maximum control. 
Exploratory plots and diagnostic plots based on Visual Predictive Checks were also extended, and bootstrap approaches used to estimate parameter uncertainty. We added several datasets to the package showcasing different types of data and analyses, with extended notebooks detailing the code and output. We also performed a simulation study based on the toenail dataset to evaluate the performance of {\sf saemix} to estimate the parameters of binary models.

{\bf Results:} {\sf saemix} showed good performance to recover the true parameter values in the simulation study, and was stable across different starting values for the parameters. The results for the different examples were in line with previous analyses on the data. Diagnostic plots and statistical criteria allowed to select appropriate models. An algorithm jointly searching for covariate and interindividual variability model was also implemented to build the covariate model and applied to a dataset containing survival data for patients with lung cancer.

{\bf Conclusions:} {\sf saemix} is a powerful and fully open-source tool for parameter estimation in NLMEM. The package has been extended to handle non-Gaussian models through a flexible interface that allows users to specify their own likelihood functions.
%{\sf saemix} is a powerful and completely open-source tool for the estimation of parameters in NLMEM. Its interface makes it easy to use and flexible, with the option to write the user's own likelihood.
% changes approved xD 

\end{abstract}

% keywords can be removed
\keywords{Non-linear mixed effect models \and SAEM algorithm \and saemix \and R package \and parameter estimation \and longitudinal data \and time-to-event \and ordinal data \and count data \and binary data}

\newpage
\section{Introduction}

Longitudinal data are collected routinely during clinical trials to monitor biomarkers, drug exposure or changes in clinical conditions. Modelling these responses can be used in routine monitoring as well as for following or predicting disease trajectories~\citep{hendrickx24}, and biomarker trajectories can be linked to survival through joint models~\citep{Desmee17}. Non-linear mixed effect models (NLMEM) are used to describe longitudinal data, as they are adapted to the hierarchical structure of repeated measurements in subjects coming from the same population. Whether set in a maximum likelihood or a Bayesian framework, specific algorithms are required to circumvent technical challenges such as computing the likelihood, which has no closed-form in NLMEM. In the maximum likelihood estimation (MLE) framework, different algorithms have been proposed which can be classified in two major categories, gradient-based algorithms minimising an approximation of (minus) the log-likelihood~\citep{lindstrom90} and EM-algorithms using the complete log-likelihood considering the individual parameters as missing data.  The Bayesian framework specifies priors on the parameters of interest to compute the posterior distribution. Software implementing these approaches include commercial software such as NONMEM~\citep{Nonmem}, Monolix~\citep{LavielleMonolix} or more recently Pumas~\citep{Pumas}, and include specialised routines to manage the specificities of pharmacokinetic/pharmacodynamic data such as dose regimens or repeated occasions. Fitting NLMEM in more general software such as R~\citep{R} is also possible. Basic installations of R come with the {\sf nlme} package as one of the fifteen recommended packages~\citep{Rnlme} implementing the first-order conditional approximation (FOCE) algorithm~\citep{Pinheiro00}. Other packages have been developed such as lme4~\citep{Rlme4Bates15}, where the likelihood is either based on a linearisation of the non-linear model function or computed using adaptive Gaussian quadrature. For Bayesian users, the {\sf rstan} package~\citep{Rstan} provides an interface to the Stan software using the Hamiltonian Monte Carlo~\citep{Stan}. {\sf saemix}~\citealp{Comets11} was the first R package implementing the stochastic approximation EM (SAEM) algorithm~\citep{Kuhn05}, an algorithm which has been proven both theoretically and practically to perform extremely well in NLMEM.

The types of outcome followed are becoming more diverse and are no longer restricted to continuous measurements such as drug concentrations or biomarker levels. There are some specialised packages in {\sf R} fitting one type of model, such as {\sf glmm} for generalised linear mixed-effect models~\citep{Rglmm}, which can be used for count or categorical data, or {\sf flexsurv} for parametric modelling of survival data~\citep{Rflexsurv}. The SAEM algorithm was extended to non-Gaussian outcome namely by Savic et al. who showed better performances compared to the algorithms based on approximations of the log-likelihood for count and categorical data~\citep{Savic09, Savic10}. In the present paper, we present version 3.2 of the {\sf saemix} package~\citep{Comets17}, extended to models for discrete and survival-type outcomes~\citep{Karimi20}. This new version, available on CRAN (Comprehensive R archive network), is a comprehensive and versatile tool to fit non-linear mixed effect models in various settings. Other noteworthy additions include improved diagnostic graphs for continuous response models using a new version of the  {\sf npde} package for R~\citep{Comets08, CometsNpde21}, as well as exploratory graphs and simulation-based diagnostics for the new types of outcome. 

%\MD{ (much better)
The objective of this paper is to describe recent extensions of the {\sf saemix} package for the analysis of non-Gaussian data in NLMEM. These new capabilities are illustrated through case studies involving binary, categorical, count and time-to-event (TTE) outcomes. We complement these examples with a simulation study assessing the performance of the extended package, based on the binary model, and provide additional results for count data in an Appendix. Further extensions enabled by the flexibility of the package are discussed in the Discussion section.

%The objective of this paper is to apply NLMEM to different types of data, through examples of modelling binary, categorical, count and time-to-event (TTE) outcomes. We complement these case-studies by evaluating the performance of {\sf saemix} in a simulation study based on the binary example, with additional results for count data given in an Appendix. In the Discussion section, we also refer to additional extensions which were implemented using the flexibility of the package.

\section{Statistical methods} \label{sec:methods}

\subsection{Models}

Detailed presentations of the non-linear mixed effects model can be found in \citep{Davidian95, Pinheiro00, LavielleMonolix}.

\paragraph{Continuous outcome:} we assume observations follow a Gaussian distribution, and we can write the following general non-linear mixed effects model to describe the $j$th observation of subject $i$, $y_{ij}\in \mathbb{R}$:
\begin{equation} 
\begin{split}
y_{ij} &=f(x_{ij},\psi_i)+ g(x_{ij},\psi_i, \sigma)\epsilon_{ij} \ \ , \ 1\leq i \leq N \ \ , \ \ 1 \leq j \leq n_i \\
\psi_i & =H(\mu,c_i,\eta_i) = H(\phi_i), \quad \eta_i \sim_{i.i.d.} {\cal N}(0,\Omega) \\ \label{eq:contNLMEM}
\epsilon_{ij} &\sim_{i.i.d.} {\cal N}(0,1) \\
\end{split}
\end{equation}
where:
\begin{itemize}
\item $N$ is the number of subjects,
\item $n_i$ is the number of observations of subject $i$,
\item $f$ is the structural model defining the behaviour of the longitudinal outcome, taking values in $\mathbb{R}$,
\item the regression variables, or design variables, ($x_{ij}$) are assumed to be known, as well as the individual covariates $c_i$,
\item for subject $i$, the vector $\psi_i=(\psi_{i,l} \, ; \, 1\leq \ell \leq n_{\psi}) \in \mathbb{R}^{n_{\psi}}$ is the vector of $n_{\psi}$ individual parameters on the scale of the model $f$, while $\phi_i$ is the vector of $n_{\psi}$ individual parameters on a normally distributed scale,
\item $\mu$ is an unknown vector of fixed effects of size $n_{\mu}$,
\item $\eta_i$ is an unknown vector of normally distributed random effects of size $n_{\eta}$.
\end{itemize}
We suppose that the $\epsilon_{ij}$ and the $\eta_{i}$ are mutually independent. We further assume that after an inverse transformation we can express the parameters $\phi_i$ as a linear combination of fixed effects $\mu$, covariates $c_i$ and random effects $\eta_i$. The link function $H$ is used to transform $\phi_i$ into the parameters entering the function $f$. The residual variance is defined by the real-valued function $g$ and some unknown parameters $\sigma$. Different error models can be used in {\sf saemix}, including constant ($g=\sigma=a$), proportional ( $g=b\,f$ and $\sigma=b$), combined ($g=\sqrt{a^2+b^2\,f^2}$ and $\sigma=(a,b)$) and exponential, where we transform the observations as $y = f \exp(a\epsilon)$ and use a constant error model.

\paragraph{Discrete outcome:} Discrete and time-to-event models are defined through the distribution for the observations, while the second-level model defining the individual parameters remains unchanged. Considering the observations $(y_{ij},\, 1 \leq j \leq n_i)$ for any individual i as a sequence of conditionally independent random variables taking their values in a fixed and finite set of nominal modalities $\{M_1, M_2,\ldots , M_K\}$, the model is completely defined by the probability mass functions $\mathbb{P}(y_{ij}=M_k | \psi_i)$ for $k=1,\ldots, K$ and $1 \leq j \leq n_i$. For a given $(i,j)$, the sum of the $K$ probabilities is $1$, so in fact only $K-1$ of them need to be defined. In the most general way possible, any model can be considered so long as it defines a probability distribution, i.e., for each $k$, $\mathbb{P}(y_{ij}=M_k | \psi_i) \in [0,1]$, and $\sum_{k=1}^{K} \mathbb{P}(y_{ij}=M_k | \psi_i) =1$. Ordinal data further assume that the categories are ordered, i.e., there exists an order $\preceq$ such that

$$M_1 \preceq M_2 \preceq \ldots \preceq M_K $$

We can think, for instance, of levels of pain $(\textrm{low} < \textrm{moderate} < \textrm{severe})$ or scores on a discrete scale, e.g., from 1 to 10. Instead of defining the probabilities of each category, it may be convenient to define the cumulative probabilities $\mathbb{P}(y_{ij} \preceq M_k | \psi_i)$ for $k=1,\ldots ,K-1$, or in the other direction: $\mathbb{P}(y_{ij} \succeq M_k | \psi_i)$ for $k=2,\ldots, K$. Any model is possible as long as it defines a probability distribution, i.e., it satisfies

$$0\leq \mathbb{P} (y_{ij} \preceq M_1|\psi_i)\leq\mathbb{P}(y_{ij} \preceq M_2|\psi_i)\leq \cdots \leq\mathbb{P}(y_{ij} \preceq M_K|\psi_i)=1 \quad \text{and }\sum_{k=1}^K \mathbb{P}(y_{ij} = M_k | \psi_i)=1$$

This definition extends easily to count data, which can take a number of possible values, possibly infinite, and it is then usual to define a probability distribution parameterised by a few parameters, such as the Poisson distribution (see case-study in section \ref{sec:case-study-count} for examples).

\paragraph{Time-to-event outcome:} Let $T_i$ denote the time to event and $\delta_i$ the indicator variable for individual $i$, with $\delta_i = 1$ if the event is observed and $\delta_i = 0$ if $T_i$ is a censoring time. The instantaneous risk of event for individual $i$ is described by the hazard function:
\begin{equation}
    h(t, \psi_i) = \lim\limits_{dt \rightarrow 0^+} \frac{Pr(t\le T_i<t+dt|T_i \ge t ; \psi_i)}{dt}
\end{equation}

A parametric hazard function allows the determination of the probability to be event-free $S$ at time $t$ by integrating the hazard until time $t$:
\begin{equation}
    S(t, \psi_i) = \exp\left(-\int_0^t h(x, \psi_i)dx\right)
\end{equation}

The likelihood of an event exactly observed at time $t$, for individual $i$, conditional on the individual parameter $\psi_i$, is the product of the survival function (=the probability of {\bf not} having experienced the event before $t$) and the hazard function (the probability to observe the event at $t$):
$$ l(t, \delta_i, \psi_i) = S(t, \psi_i) \; h(t, \psi_i)^{\mathbb{1}_{\delta_i}=1} =  S(t, \psi_i) \; h(t, \psi_i)^{\delta_i}. $$

\subsection{The SAEM algorithm}

%\Eco{TODO: check notation consistency with main paper}

The SAEM algorithm was first proposed by~\citep{Kuhn05}. We give a brief description below, referring the reader to eg~\citep{LavielleMonolix, Comets17} or to the documentation of {\sf saemix} (\url{https://github.com/iame-researchCenter/saemix/blob/fcf6231d6ee8f89af01cc1a81642472560c7eb0e/docsaem.pdf)} for more details. 

%Refer to documentation~\citep{Kuhn05, Savic09, Savic10} + addition by Belhal~\citep{Karimi20}

\subsection*{Estimation of the population parameters}

The SAEM algorithm computes the maximum likelihood estimator of the unknown set of parameters $\theta=(\mu,\Omega,\sigma)$, that is, the parameters $\theta$ maximising the log-likelihood: 
\begin{equation}
LL(\theta | y) = \sum_{i=1}^N \log \left( p(y_i | \theta) \right) 
=  \sum_{i=1}^N \log \left( \int p(y_i | \theta, \psi_i) p(\psi_i | \theta) d\psi_i  \right).  \label{eq:likeNLMEM}
\end{equation}
Because the integral is intractable when $f$ is non-linear, EM algorithms iterate between an E-step, which consists at iteration $k$ in computing the conditional expectation of the complete log-likelihood 
\begin{equation}
Q_k(\theta)= \sum_{i=1}^N \mathbb{E}{[\log p(y_i, \psi_i |\theta) | y_i,\theta_{k-1}}]
\label{eq:Q-quantity}
\end{equation}
and an M-step which updates $\theta$'s estimation with the value $\theta_{k}$ that maximises $Q_k(\theta)$. Following \citep{Dempster77,Wu83}, the EM sequence $(\theta_k)$ converges to a stationary point of the observed likelihood under general regularity conditions (see also Appendix A). When the regression function $f$ does not linearly depend on the random  effects, the E-step cannot be performed in a closed-form and is replaced by a stochastic procedure. At iteration $k$ of SAEM:
\begin{itemize}
\item {\em Simulation-step} : draw $\psi_i^{(k)}$ from the conditional distribution  $p(\cdot|y_i;\theta_k)$, $1 \leq i \leq N$.
\item {\em Stochastic approximation} : update $\tilde{Q}_k(\theta)$, which approximates \eqref{eq:Q-quantity}, according to
\begin{equation}
\tilde{Q}_k(\theta) = \tilde{Q}_{k-1}(\theta) + \gamma_k ( \log p(y_i, \psi_i^{(k)} |\theta) - \tilde{Q}_{k-1}(\theta) )
\end{equation}
where $(\gamma_k)_k$ is a sequence of step sizes in $[0,1]$, decreasing towards $0$ and such that $\sum_k \gamma_k = \infty$ and $\sum_k \gamma_k^2 < \infty$.
\item {\em Maximisation-step} : update $\theta_k$ according to 
$$\theta_{k+1}={\rm Arg}\max_{\theta} \tilde{Q}_k(\theta)$$
\end{itemize}

It is shown in \citep{Delyon99} that SAEM converges to a maximum (local or global) of the likelihood of the observations under very general conditions.

In the simulation step, individual parameters $\psi^{(k)} = \{\psi^{(k)}_i, \ i=1,\dots,N\}$ cannot be drawn directly from their conditional distribution.  Instead, they are simulated using Markov Chain Monte Carlo (MCMC) methods, 
most commonly the Metropolis–Hastings (MH) algorithm.
%In the simulation step, individual parameters $\psi^{(k)} = \{(\psi^{(k)}_i), \ i=1,...,N \}$ are simulated from their conditional distribution with a Metropolis-Hasting (MH) algorithm. 
Successive and different transition kernel and associated proposals may be used (see e.g. \citep{LavielleMonolix} for details).
%\begin{itemize}
%    \item the prior distribution of the individual parameter at the previous step,
%    \item a multidimensional random walk whose parameters depend on the estimates from the previous step, 
%    \item and finally, a succession of unidimensional Gaussian random walks updating each component of the individual parameter vector successively
%\end{itemize}
%These successive proposals are centered at the current state with a diagonal variance-covariance matrix. The latter is adjusted at each iteration with a multiplicative factor.

In the maximisation step, the population parameters are updated through sufficient statistics of the model.
%, with an additional minimisation step for certain model structures (combined error model, parameters without IIV, ...).
For certain model structures (combined error model, parameters without IIV, ...) an additional minimisation step may be performed, depending on whether the model belongs to the curved exponential family or not~\citep{LavielleMonolix}.
\paragraph{\itshape Iteration phases:} For effective exploration of the parameter space, the algorithm is structured in two phases.
%The algorithm consists in 2 phases. 
In the first $K_1$ iterations (exploration phase), we initialise the algorithm with a burn-in sequence of 5 iterations where samples are discarded after each iteration, and then let the algorithm explore the parameter space by setting $\gamma_k=1$. In the smoothing phase ($K_2$ iterations), we set $\gamma_k = 1/(k-K_1+1)$ to ensure almost sure convergence of the algorithm to the MLE. By default, simulated annealing is implemented during the first half of the exploration phase to increase the variance of the proposals.

In order to improve overall convergence, it is possible to run $L$ Markov Chains instead of only one, which means drawing sequences $ \psi^{(k,1)}, ...,  \psi^{(k,L)}$ in the simulation step, and then combining the stochastic and the Monte-Carlo approximations:
$$\tilde{Q}_k(\theta) = \tilde{Q}_{k-1}(\theta) + \gamma_k \left( \frac{1}{L} \sum_{\ell=1}^L \log p(y_i, \psi_i^{(k, \ell)} |\theta) - \tilde{Q}_{k-1}(\theta)\right)$$

\paragraph{\itshape f-SAEM:} the fast-SAEM algorithm was added to {\sf saemix} in version 3.0 to accelerate the convergence of the MLE estimation~\citep{Karimi20}. A fourth kernel is added in the simulation step, using a proposal approximating the conditional distribution of the individual parameters. For continuous data models, the covariance of the proposal is obtained by a first-order linearisation of the model while for non Gaussian models, it is derived using a Laplace approximation.

\subsection*{Estimation of the likelihood}

The likelihood of the non-linear mixed effects model defined in~\eqref{eq:likeNLMEM} cannot be computed in a closed-form. A first option is to approximate this likelihood  by the likelihood of the Gaussian model deduced from the non-linear mixed
effects model after linearisation of the function $f$ around the predictions of the individual parameters $(\phi_i, 1\leq i \leq N) $.

Alternatively, the likelihood can be estimated without any approximation to the model using Monte-Carlo or quadrature approaches. In {\sf saemix}, stochastic integration is implemented through an importance sampling approach, while numerical integration involve Gauss-Hermite quadrature methods. For both approaches, the quality of the computation depends on the estimates of the conditional mean and variances of the individual distributions.

\subsection*{Estimation of the uncertainty}

For continuous outcome models, {\sf saemix} provides by default estimates of the uncertainty of parameter estimates through an asymptotic approximation based on the expected Fisher Information Matrix (FIM). The FIM is computed using a linear approximation of the model function $f$ (also called the FOCE method). Because this approximation is poor for discrete outcome models, the current version of {\sf saemix}  does not compute the linearised FIM for these models. Bootstrap approaches have been proposed for NLMEM~\citep{Comets21} and are implemented in the package. Alternative methods based on MH algorithms are being investigated~\citep{Guhl22, Delattre23, LavalleyPAGE23}.

\subsection*{Estimation of the individual parameters}

When the parameters of the model have been estimated, the individual parameters $(\psi_i)$ are estimated via their conditional distributions. We first estimate the individual normally distributed parameters $(\phi_i)$ and derive the estimates of $(\psi_i)$ using the transformation $\psi_i= H(\phi_i)$. The conditional distributions are estimated through the MCMC procedure used in the simulation step, using the population estimates as priors and iterating until a prespecified tolerance is reached. Individual parameters can then be derived as the mode (Maximum A Posteriori) or the mean of the conditional distribution, and the conditional standard deviation provides an estimate of their uncertainty.

\section{saemix 3}

\paragraph{The {\sf \bfseries saemix} package:} the first version of the package was uploaded to CRAN in 2011~\citep{Comets11} and was an {\sf R} version of the original Matlab code of Monolix~\citep{LavielleMonolix}. Since then regular updates were released with new features to facilitate model building and evaluation, and a paper showing the good performance of {\sf saemix} in comparison to {\sf nlme} and {\sf nlmer} was published in 2017~\citep{Comets17}. The extension to non-Gaussian outcomes was included in version 3.0, with the current version at the date of submission being 3.5 (February 2026). Diagnostic graphs for NLMEM are useful tools to diagnose model deficiencies~\citep{Nguyen17} and we changed the graphs produced by {\sf saemix} to make use of our companion package {\sf npde}~\citep{CometsNpde21} when we made the switch to {\sf ggplot2}~\citep{Rggplot2} in version 3.0. 
%In addition, Dr Delattre implemented a Bayesian Information Criterion (BIC) adapted to NLMEM in {\sf saemix}~\citep{Delattre14} as well as a stepwise algorithm for automated model building of the covariate model and the variability structure~\citep{Delattre20}.
In addition, a Bayesian Information Criterion (BIC) adapted to NLMEM~\citep{Delattre14} as well as a stepwise algorithm for automated model building of the covariate model and the variability structure~\citep{Delattre20} were implemented in {\sf saemix}.
Finally, we added the option to switch to the fast-SAEM algorithm proposed by~\citep{Karimi20} for the first iterations of the estimation process to accelerate convergence.

A detailed run-through the main functions in the package is available in the documentation, which can be downloaded on a public repository on github (\url{https://github.com/iame-researchCenter/saemix}). For the sake of simplicity in the present paper, we will defer the code itself to the notebooks accompanying the paper. The development version of the package is available on the development github repository (\url{https://github.com/saemixdevelopment/saemixextension}), and all the code, as well as additional analyses, is made available through specific notebooks for each type of outcome, found in the folder (\url{https://github.com/saemixdevelopment/saemixextension/blob/master/paperSaemix3/}). 

\paragraph{Package extensions:} To fit discrete outcome models in {\sf saemix}, we write the model function computing the log-likelihood of the observations. An important difference with modelling longitudinal outcomes is that for this computation we need to pass the observed values as predictors and therefore include them in the {\sf saemixData} object. 

Plots to explore the data depending on the type of outcome are available through the {\sf plotDiscreteData()} function, specifying the outcome type as binary, categorical, count or event. Many functions in the package will work similarly, such as the extraction of results after a fit or summary functions, while others will adjust to the outcome. For example, the prediction functions will predict the log-likelihood of the observations according to the fitted model. Another difference with longitudinal outcomes is that because the user is free to write the log-likelihood directly, there is no associated simulation function by default. To obtain simulations under the model, the user can add an additional element of the model object to specify a simulation function based on the underlying probability model for the observations. Doing this will allow {\sf saemix} to produce simulated datasets under the model, and some diagnostics will then be available through the {\sf discreteVPC()} function.

Finally, standard error of estimations can no longer be obtained through the Fisher Information Matrix as implemented currently, since the approximation used in {\sf saemix} is a first-order approximation of the FIM which is known to behave poorly with discrete data outcomes. Instead, bootstrap approaches have been implemented in {\sf saemix} which can be applied to these types of models~\citep{Comets21}, noting that residual-based bootstrap may require defining a simulation function, same as for diagnostics.

\paragraph{Examples of use:} Several examples are used to illustrate typical modelling workflows, including data exploration and formatting, model definition, parameter estimation, standard errors of estimation, model comparison through statistical criteria, and model diagnostics. Model diagnostics for longitudinal outcomes are now provided through the {\sf npde} package~\citep{CometsNpde21}, using simulations from the models fit in {\sf saemix}. Examples for discrete outcomes have been added in the current extension, with the corresponding datasets available and documented in the package: binary data ({\sf toenail.saemix}), categorical data ({\sf knee.saemix}), count data ({\sf rapi.saemix}) and time-to-event data ({\sf lung.saemix}).

%\newpage
\section{Non-Gaussian data models using the SAEM algorithm}

\subsection{Binary data}

\subsubsection{Case study: the toenail data} % (from paper on npd for categorical data, submitted)

\hskip 18pt The {\sf toenail.saemix} dataset in the {\sf saemix} package contains binary data from a randomised clinical trial comparing two treatments for fungal toenail infection, available in {\sf R} as the {\sf Toenail} dataset in the {\sf R} package {\sf prLogistic}~\citep{Rprlogistic}. Data are from \citep{debacker_toenail}, a multi-center randomised comparison  of two oral treatments (A and B) for toenail infection, where 294 patients were measured at seven visits (baseline=week 0, weeks 4, 8, 12, 24, 36, and 48), comprising a total of 1908 measurements. The primary end point was the presence of toenail infection and the outcome of interest is the binary variable {\it onycholysis} which indicates the degree of separation of the nail plate from the nail-bed (none or mild versus moderate or severe).  

\begin{figure}[!h]
    \centering
    \includegraphics[scale=0.4]{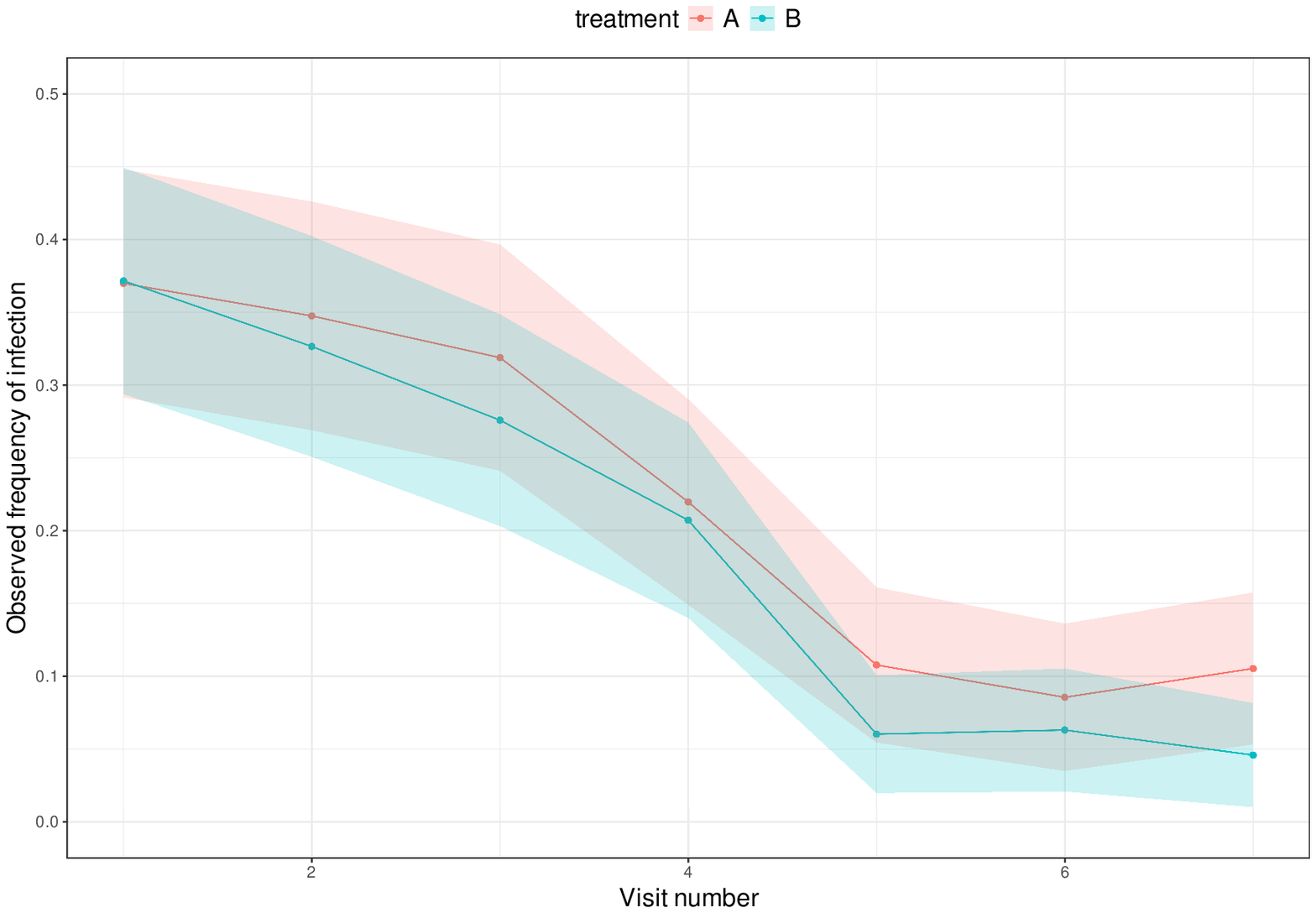}
    \caption{\label{fig:toenailData} Time-course of the proportion of subjects with toenail infection at each visit, stratified by treatment.}
\end{figure}

%Several analyses have been made, as in~\citep{lesaffre2001effect,lin2011goodness} and the logistic random effect model proposed by \citep{hedeker1994random} is considered in this work.  % checked, Hedeker doesn't use the toenail data... (but develops the mixor software later used to fit that data)

Several analyses have been made, as in~\citep{lesaffre2001effect,lin2011goodness} and the logistic random effect model proposed by \citep{lesaffre2001effect} is considered in this work. This model includes a random intercept ($\theta_{1,i}$), a time effect ($\theta_2$) and treatment (A or B) ($\beta$) as a covariate affecting the slope $\theta_2$. We considered the interaction term between time and treatment but no treatment on the intercept, which shouldn't be different between arms due to the randomisation process. The time course was assumed to be similar across subjects as no significant interindividual variability was found when testing different models. {\sf saemix} was used to fit a linear model on the logit scale (${\rm logit}(p) = \ln \left( \frac{p}{1-p} \right)$):

%$$ {\rm logit}(P(y_{ij}=1)) = \theta_{1,i} + (\theta_{2} + \beta \mathbb{1}_{trt=B}) t_{ij} $$

\begin{align}
y_{ij} \mid \theta_{1,i} &\sim \mathrm{Bernoulli}(p_{ij}), \nonumber\\
\mathrm{logit}(p_{ij}) &= \theta_{1,i} + (\theta_2 + \beta \mathbb{1}(\mathrm{trt}_i = B))\, t_{ij}, \label{eq:bernoulli} \\
\theta_{1,i} &\sim \mathcal{N}(\theta_1, \omega_1^2). \nonumber
\end{align}

Table~\ref{tab:toenailParameters} shows the parameter estimated with this model, as well as the standard errors of estimation (SE) estimated using two bootstrap approaches, the classical case bootstrap with resampling at the level of the individual and the non-parametric conditional bootstrap proposed for non-linear mixed effect models~\citep{Comets21}. The SE for all parameters except $\theta_1$ were lower when using the conditional bootstrap, and a consequence is that treatment effect which was not significant with the SE estimated by case bootstrap became (slightly) significantly different from 0 for the conditional bootstrap. While in this case we can't determine which of the methods is correct, the simulation study performed in the next section suggests the conditional bootstrap may be closer to the empirical SE in this instance.

\begin{table}[!h]
\begin{center}    
    \begin{tabular}{cccccc}
         \hline
         Parameter & Estimate & \multicolumn{2}{c}{Case bootstrap} & \multicolumn{2}{c}{Conditional bootstrap} \\
         && RSE\% & CI & RSE\% & CI\\
         \hline
         $\theta_1$ (-) & -1.71 & 21 & [-2.44, -1.08] & 18 & [-2.20, -0.98]\\
         $\theta_2$ (d$^{-1}$) & -0.39 & 18 & [-0.56, -0.28] & 10 & [-0.49, -0.32]\\
         $\beta$ (-) & -0.15 & 87 & [-0.44, 0.07] & 47 & [-0.29, -0.02]\\
         $\omega_1$ (-) & 4.02 & 13 & [ 3.22, 5.31]  & 9 & [ 3.16, 4.64] \\
         \hline \\
    \end{tabular}
    \caption{Parameter estimates for the logistic model adjusted to the toenail data using {\sf saemix}. The SE and CI were estimated using case and conditional bootstrap. \label{tab:toenailParameters}}
\end{center}
\end{table}

% Estimates obtained with Monolix 
% higher SE estimated for beta and theta2 with the case bootstrap
% on the other hand the SE reported by Monolix very close to those obtained with the conditional bootstrap... (but treatment effect becomse significant)
% S.E.	R.S.E.(%)
% th1_pop	-1.74	0.34	19.5
% th2_pop	-0.38	0.039	10.3
% beta_th2_treatment_1	-0.15	0.061	40.0
% omega_th1	4.01	0.38	9.38

\subsubsection{Performance of {\sf saemix}}

In this section, we evaluate the performance of the package to estimate the population parameters in binary models, using the {\sf toenail} dataset as a template for a simulation study.

\paragraph{Simulation settings:} the simulated times were set at the original 7 times included in the protocol, that is baseline, 1, 2, 3, 5.5, 8 and 11 months when converted from weeks to months (rounded). The settings included 274 subjects, with 137 in each treatment group.

Two designs were simulated. In the first, mimicking the {\sf toenail} dataset, we used the parameters reported in Table~\ref{tab:toenailParameters}  (rounded to 2 digits). In the second, we assumed variability on both parameters, i.e. $\theta_{1,i} \sim \mathcal{N}(\theta_1, \omega_1^2)$ and $\theta_2$ in~\eqref{eq:bernoulli} replaced by $\theta_{2,i} \sim \mathcal{N}(\theta_2, \omega_2^2)$ with $\omega_1 = 1$, $\omega_2=0.2$. The designs were checked for identifiability by computing the predicted Fisher Information Matrix (FIM) using the MC/AGQ approach proposed by~\citep{Ueckert17}. In both scenarios, we simulated 200 datasets and estimated the population parameters using {\sf saemix}.  To assess the stability of the estimates, we used 3 settings, either starting from the true value of the parameters, from a rough population estimate based on the frequency of toenail infection at baseline and at the last visit, without treatment effect and a smaller initial value for the variability ($\theta_{1, 0}=-0.5$, $\theta_{2, 0}=-0.19$, $\beta_{0}=0$, $\omega_{1, 0}=1$), or from values of 0 for the parameters and 4 for the variability. In all cases, 10 chains were used, with simulated annealing active for the first $K_1/2$ iterations (see documentation for details of the algorithm).

%ie around 50\% IIV for both parameters) 

% /home/eco/work/saemix/discreteEval (run in August 2022)

\paragraph{Evaluation:} We computed the relative bias as the mean of the relative estimation errors, and the relative root mean square error (RMSE) as their standard deviation, expressed in percentage:
\begin{equation}
\begin{split}
RB(\hat{\theta}) &= \dfrac{1}{S} \sum_{s=1}^S \dfrac{\hat{\theta}_s - \theta^0}{\theta^0} \times 100 \\
RRMSE(\hat{\theta}) &=  \sqrt{\dfrac{1}{S} \sum_{s=1}^S \left( \dfrac{\hat{\theta}_s - \theta^0}{\theta^0} \right)^2 } \; \times 100 \\
\end{split}
\end{equation}
where $\theta^0$ refers to the true value of the parameter, $\hat{\theta}_s$ to its estimation at the s-th simulation, and $S$ to the number of simulated datasets.

\begin{figure}[!h]
    \centering
    \includegraphics[scale=0.4]{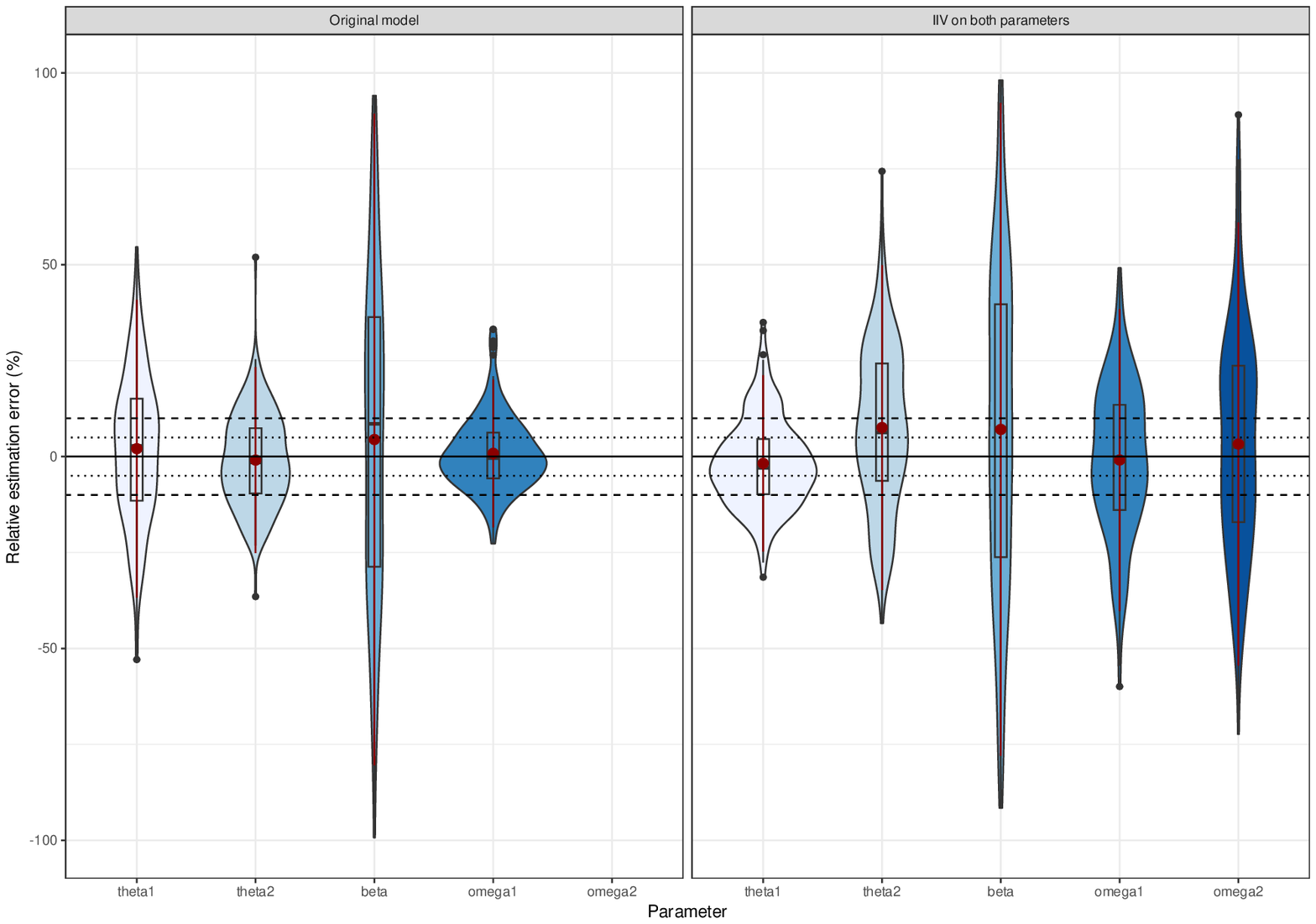}
    \caption{\label{fig:binSimulREE} Relative estimation errors for the parameters in the original design (left) and the design with variability on both parameters (right). Dashed lines delineate absolute relative biases within 10\% and dotted lines denote biases within 5\%. The red insert highlights the interval corresponding to the mean (red dot) plus or minus 2 standard deviations. The graph was trimmed to +/-100, omitting 26 values for $\beta$. % (using the function {\sf mean\_sdl} as in input to {\sf stat\_summary} in {\sf ggplot2}).
    }
\end{figure}

\paragraph{Results: } Figure~\ref{fig:binSimulREE} shows a violin plot of relative estimation errors (REE) for the different parameters in the two designs. All settings gave the same results, suggesting the runs were stable with 10 chains, so for clarity we only show the plots for the setting with the true parameters as starting values (full results are in Appendix). In the original design, the bias was less than 5\% for all parameters, as shown in table~\ref{tab:binSimulREE} in the appendix, and the RRMSE was between 10 and 20\% for all parameters, except $\beta$ which shows more variability in the estimates (50\%). When we set variability for both parameters in the model, we observed a bias around 12\% for $\beta$ and 8\% on $\theta_2$, with increased variability between estimates in the different replicates especially for $\beta$ and the variability of the slope $\omega_2$, which were accordingly the two parameters with the highest RRMSE. The estimated RRMSE was close to the values predicted by the FIM given the design and parameters, except for $\beta$ in the second scenario where it was smaller (56\%) than predicted (112\%). Finally, we can compare the relative RMSE for the first scenario, given in Table~\ref{tab:binSimulREE} of the Appendix, to the values estimated by bootstrap for the same model in the real data example (table~\ref{tab:toenailParameters}) to find that conditional bootstrap estimates were very close to the empirical estimates obtained over the 200 simulations and more accurate in this case than case bootstrap estimates.

% Expected FIM
% Orig -24.86142 -11.63633 -38.29206  20.39756  (omega2 ?)
% IIV -10.70774  -20.99451 -112.04915   36.80376   65.15504 

%\clearpage
%\newpage

\subsection{Categorical data}

% \Eco{Points of interest:}
% \begin{itemize}
% \item model for categorical data
% \item data exploration and model diagnostics
% \item bootstrap to obtain SE
% \item results (compared to vignette if exists ? other analysis ?)
% \end{itemize}

\paragraph{Data:} Next, we illustrate the use of {\sf saemix} to model categorical data. The {\sf knee.saemix} dataset represents pain scores recorded in a clinical study in 127 patients with sport related injuries treated with two different therapies. The pain occuring during knee movement was observed after 3,7 and 10 days of treatment. It was taken from the {\sf catdata} package in R~\citep{Rcatdata} (dataset {\sf knee}) and reformatted for {\sf saemix}, adding a time column representing the day of the measurement (with 0 being the baseline value), so that each observation corresponds to a different line in the dataset. Treatment was recoded as 0/1 (placebo/treatment), gender as 0/1 (male/female) and {\sf Age2} represents the square of centered {\sf Age}. 

We can visualise the evolution of the proportion of each score over time using the {\sf plotDiscreteData()} function with the outcome set to {\sf categorical}, stratifying by treatment (Figure~\ref{fig:dataKnee} in Appendix). In both treatment groups, the probability of lower scores (1 or 2) increases with time while the probability to obtain a high (4 or 5) pain score decreases, suggesting a recovery, while the treatment group itself seems to impact more the median than the extreme pain scores.

\paragraph{Models:} The dataset {\sf knee} is part of the datasets analysed in~\citep{Tutz11} with various methods described in the vignettes in the documentation of the {\sf catdata} package, but mainly as logistic regression on the response after 10 days, or as mixed binary regression after dichotomising the response. Here, we fit a proportional odds model to the full data.  The probability $p_{ij}=P(Y_{ij}=1 | \psi_{i})$ associated with an event $Y_{ij}$ at time $t_{ij}$ is given by the following equation for the logit:
\begin{equation}
\begin{split}
\mathrm{logit}(P(Y_{ij} = 1 | \psi_i)) &= \theta_{1,i} + \alpha_{i} t_{ij}, \\
\mathrm{logit}(P(Y_{ij} = 2 | \psi_i)) &= \theta_{1,i} + \theta_{2,i}, \\
\mathrm{logit}(P(Y_{ij} = 3 | \psi_i)) &= \theta_{1,i} + \theta_{2,i} + \theta_{3,i}, \\
\mathrm{logit}(P(Y_{ij} = 4 | \psi_i)) &= \theta_{1,i} + \theta_{2,i} + \theta_{3,i} +\theta_{4,i},\\
P(Y_{ij} = 4 | \psi_i) &= 1 - \sum_{k=1}^4 P(Y_{ij} = k | \psi_i),\\
\end{split}
\end{equation}
where $\theta_{1,i}$ is the logit-transformed probability of a pain score of 1 at time 0, the other $\theta_{k,i}$ parameters represent an incremental risk to move into the higher pain category, and $\alpha_i$ represents an effect of time for individual $i$.

We ran the SAEM algorithm with 10 chains and 600 and 100 iterations respectively in the exploration and smoothing phases. We first fit a model without covariates, with variability on $\theta_{1,i}$ and $\alpha_i$ only, referred to as the base model hereafter. $\theta_{1,i}$ was assumed to have a normal distribution, $\theta_{1,i} \sim \mathcal{N}(\theta_1, \omega_{\theta_1}^2)$, while $\alpha_i$ was assumed to have a log-normal distribution, $\alpha_i \sim \mathrm{LN}(\alpha, \omega_{\alpha}^2)$, to reflect the increased probability of a lower pain score as time passes. The remaining parameters were treated as fixed (without variability) and were constrained to be positive, specified by {$\theta_{k,i} = \exp(\theta_k)$ for all $i=1,\ldots,N$ and $k=2,3,4$, to preserve the ordinal structure of the data}. In a second step, we used the stepwise algorithm implemented by~\citep{Delattre20} based on the BICc associated with non-linear mixed effect models~\citep{Delattre14} to simultaneously explore the variability structure and the covariate model. In this second step, the parameters $\theta_{k,i}$, $k=2,3,4$, were assumed to follow log-normal distributions, $\theta_{k,i} \sim \mathrm{LN}(\theta_k,\omega_{\theta_k}^2)$. The resulting model from this stepwise approach is called the covariate model. Diagnostics were obtained for the base and covariate models.

\paragraph{Results:}  Table~\ref{tab:kneeEstimates} reports the estimated parameters for the base and covariate models. As previously, we assessed parameters uncertainty using bootstrap approaches, reporting the bootstrap confidence intervals obtained using the conditional bootstrap approach as it gave better results in the simulation for binary responses.
\begin{table}[!h]
\begin{center}    
    \begin{tabular}{lcccc}
\hline
  Parameter & \multicolumn{2}{c}{Base model} & \multicolumn{2}{c}{Covariate model} \\
  & Estimate & Bootstrap CI & Estimate & Bootstrap CI \\ 
  \hline
$\theta_1$ & -15.2 & [-19.0, -10.7] & -23.3 & [-28.1, -13.6] \\ 
  $\beta_{Age^2,\theta_1}$ & - & - &   0.06 & [  0.03, 0.09] \\ 
  $\theta_2$ &   6.5 & [4.6, 7.7] &   3.5 & [  1.3, 4.8] \\ 
  $\beta_{Trt,\theta_2}$ & - & - &   1.0 & [  0.43, 1.6] \\ 
  $\theta_3$ &   8.5 & [6.6, 10.6] &   8.5 & [  5.8, 11.2] \\ 
  $\theta_4$ &  12.5 & [9.7, 17.0] &  16.4 & [  9.4, 20.7] \\ 
  $\alpha$ &   0.87 & [0.6, 1.0] &   0.88 & [  0.55, 1.0] \\ 
  $\beta_{Trt,\alpha}$ & - & - &   0.56 & [  0.21, 0.83] \\ 
  $\omega_{\theta_1}$ & 13.8 & [10.5, 17.0] & 15.6 & [ 9.3, 17.8] \\ 
  $\omega_{\theta_2}$ & - & - &   0.43 & [0.01 , 0.32] \\ 
  $\omega_{\theta_3}$ & - & - &   0.74 & [ 0.25, 0.89] \\ 
  $\omega_{\theta_4}$ & - & - &   0.69 & [ 0.18, 1.1] \\ 
  $\omega_{\alpha}$ & 0.74 & [0.55, 0.95] & - & - \\ 
  \hline \\
    \end{tabular}
    \caption{Parameter estimates for the proportional odds models without and with covariates. The RSE and CI were obtained using the conditional bootstrap\label{tab:kneeEstimates}.}
\end{center}
\end{table}

\begin{figure}[!h]
    \centering
    \includegraphics[scale=0.4]{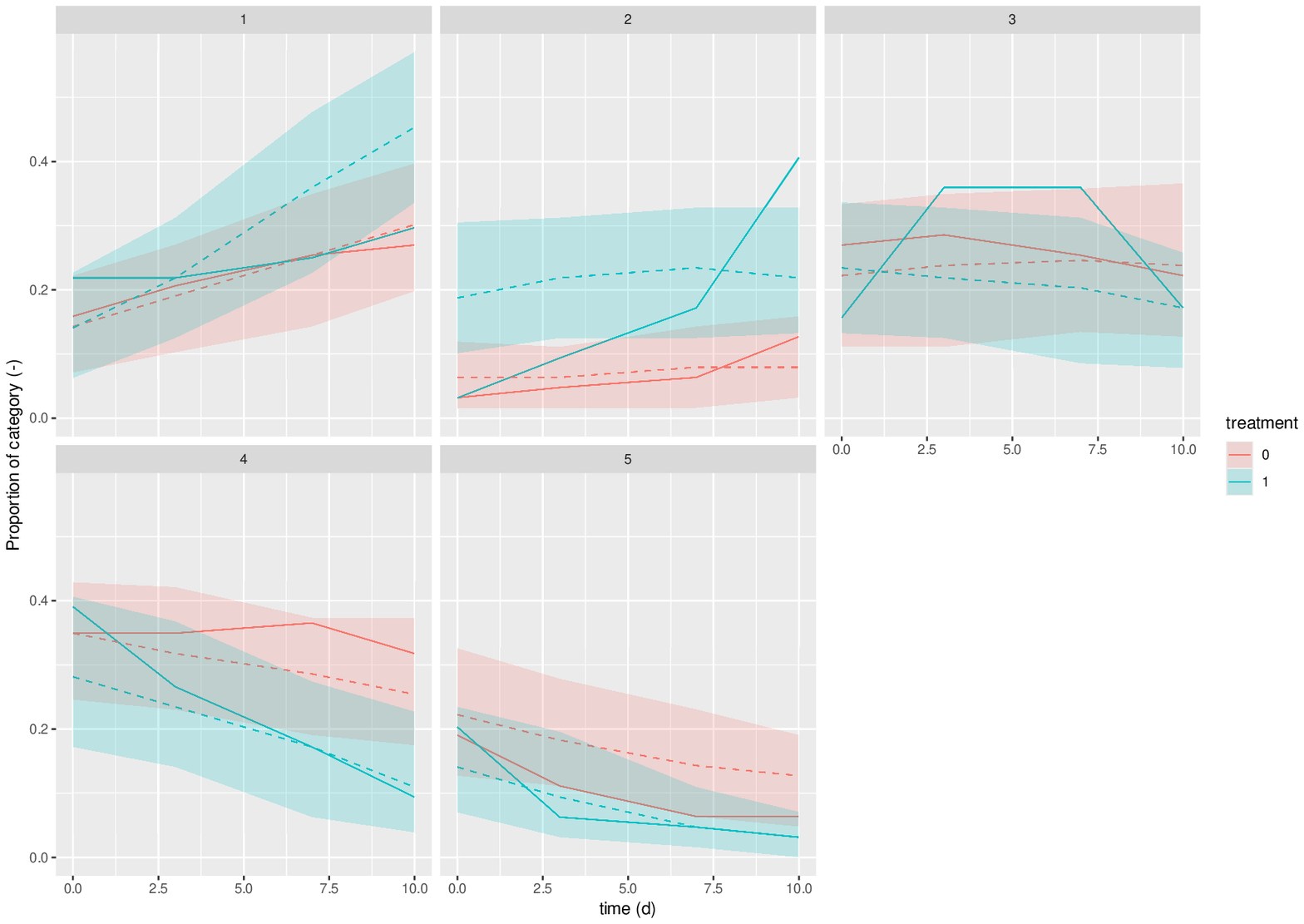}
    \caption{\label{fig:kneeVPC} Visual Predictive Check for each value of the score in the proportional odds model with covariates, stratified by treatment. VPC are produced by simulating descriptive statistics under the model and design of the original dataset and comparing them with the observed values. Here, we simulated the scores for 1000 datasets under the same design, then computed the 95\% prediction intervals of the proportion of each scores over time in each treatment group. We then overlayed the observed proportion.}
\end{figure}

Figure~\ref{fig:kneeVPC} displays Visual Predictive Checks (VPC) stratified by treatment. The figure shows some model misspecification, especially in the intermediate pain scores, as well as a tendency to overestimate the improvement, driven by the increase in the lowest pain score. This suggests treatment effect is not properly accounted for the three lowest pain categories and across time. Although gender was not retained in the covariate model, VPC stratified on gender seem to indicate different distributions of the pain scores between men and women, especially in the first and third categories (no or middle pain levels, Figure~\ref{fig:kneeVPCSex} in Appendix C). We can also look at the VPC for the median score in each treatment group to find that the model tends to underpredict the pain scores, especially in the group receiving therapy 1 (Figure~\ref{fig:kneeVPCmedian} in Appendix C). Taken together, these results suggest that the proportional odds model may not be adequate to describe this data, and we could investigate alternative models such as differential odds models or bounded integer models. 

% hard to find a previous longitudinal analysis of the data to compare to :-/

% REMOVED: performance evaluation for categorical data
% simulation study based on an analysis of WHO-score in Discovery so would be a bit difficult to justify
% 3 categories: good results for saemix, F-saem and Monolix
% 7 categories: Monolix shows good results with no bias, F-saem also good, similar RSE, but saemix on this example has bias (not sure about the scale, maybe less than 5% bias or could be 500%...) and RMSE twice as high as the other two (but again not sure if the scale is indeed %, then only 8% so not catastrophic) $\Rightarrow$ need to investigate further and set up simulations.

\newpage
\subsection{Count data} \label{sec:case-study-count}

% RAPI data, re-analysis with saemix as in notebook. \Eco{Points of interest:}
% \begin{itemize}
% \item covariate effects, model comparisons
% \item reproduce previous analysis
% \item data exploration and model diagnostics
% \item bootstrap to obtain SE
% \end{itemize}

\paragraph{Data:} The {\sf rapi.saemix} dataset in the {\sf saemix} package contains count data kindly made available by David Atkins (University of Washington) in his tutorial on modelling count data~\citep{Atkins13}. The data come from a randomised controlled trial assessing the effectiveness of web-based personalised normative feedback intervention on alcohol consumption~\citep{Neighbors10a, Neighbors10b}. The {\sf rapi.saemix} dataset records alcohol-related problems, as measured by the Rutgers Alcohol Problem Index (RAPI)~\citep{White89}, in freshmen at risk for heavy drinking behaviours. Students were asked to report every six months the number of alcohol-related problems, and the dataset includes over 3500 repeated measures of these counts in 818 subjects. Interesting features of this dataset are first, the longitudinal aspect which allows to evaluate changes over time, and second, the shape of the distribution of counts. A histogram of the data shows a large proportion of data points at zero, indicating individuals or occasions without problems. In Figure~\ref{fig:dataRapi} (Appendix D), we show the evolution of the proportion of each count over the two years of the study. This dataset was used in~\citep{Atkins13} as a tutorial in mixed effects count regression using the {\sf glmer()} function from the {\sf lme4} package~\citep{Rlme4Bates15}. Here, we show how to use {\sf saemix} to fit the main models used in the analysis of count data. 

\paragraph{Models:} The first model that comes to mind when considering count data is the \textbf{Poisson model}, considering the logarithm of the mean of the number of events ($\lambda_i$) to be a linear function of time:
%\begin{equation*}
%\begin{split}
%P(Y=n) &= \frac{\lambda^n \; e^{-\lambda}}{n!} \\
%\log(\lambda) & = \alpha_0 + \alpha_1 t \\
%\end{split}
%\end{equation*}
\begin{equation*}
\begin{split}
P(Y_{ij}=n|\alpha_{0,i}, \alpha_{1,i}) &= \frac{\lambda_i(t_{ij})^n \; e^{-\lambda_i(t_{ij})}}{n!} \\
\end{split}
\end{equation*}
where
\begin{equation}
\log(\lambda_i(t)) = \alpha_{0,i} + \alpha_{1,i} t 
\label{eq:lambda}
\end{equation}
Following~\citep{Atkins13}, we include a gender effect on both parameters:
\begin{equation}
\alpha_{k,i} = \alpha_k + \beta_{male,\alpha_k}\mathbb{1}_{(\mathrm{gender_i} = \mathrm{male})} + \eta_{k,i} \; , \; \eta_{k,i} \sim \mathcal{N}(0,\omega_{\alpha_k}^2) \; , \; k=0,1
\label{eq:lambda_pop}
\end{equation}

The VPC for this model, shown in Appendix D (Figure~\ref{fig:rapiPoissonVPC}), highlight that the Poisson model isn't flexible enough to accommodate the large number of zero's in the data. 

Two other models that can be applied are the zero-inflated Poisson (ZIP) model and the hurdle model. The \textbf{ZIP-model} uses a mixture to account for the proportion of zero's:
\begin{equation*}
\begin{split}
P(Y_{ij}=0|\alpha_{0,i}, \alpha_{1,i}) &= p_0 + (1-p_0) e^{-\lambda_i(t_{ij})} \\
P(Y_{ij}=n | Y_{ij}>0, \alpha_{0,i}, \alpha_{1,i}) &= \frac{\lambda_i(t_{ij})^n \; e^{-\lambda_i(t_{ij})}}{n!} \\
\end{split}
\end{equation*}
with $\lambda_i(t)$ defined as in~\eqref{eq:lambda}. No variability is set on $p_0$ which represents a proportion at the level of the population and its value is constrained to be in $[0,1]$. Individual parameters are defined as in~\eqref{eq:lambda_pop}.
%and we use a logit-normal distribution for this parameter to ensure it remains between 0 and 1. 
The {\bf hurdle model} on the other hand is a combination of two models fit on separate sets of data: a binary logistic regression to determine the probability of zero counts (fit on the entire dataset modelled as one or more events versus none) and a truncated Poisson model to describe non-zero counts (fit only on strictly positive counts). In this model, we fit:
\begin{equation*}
\begin{split}
\text{Hurdle 0: } & \text{binary logistic regression} \\ 
P(Y_{ij}>0|\alpha_{0,i}, \alpha_{1,i}) &= e^{\alpha_{0,i}+\alpha_{1,i} \; t_{ij}}/(1+e^{\alpha_{0,i}+\alpha_{1,i} \; t_{ij}}) \\
\text{Hurdle 1: } & \text{truncated Poisson model}\\ 
P(Y_{ij}=n | Y_{ij}>0, \lambda_i) &= \frac{\lambda_i(t_{ij})^n \; e^{-\lambda_i(t_{ij})}}{(1-e^{-\lambda_i(t_{ij})})\;n!}, \\
\end{split}
\end{equation*}
where $\lambda_i(t)$ defined as in \eqref{eq:lambda}-\eqref{eq:lambda_pop}. The main difference between the two models is that the hurdle model separates the zero from the non-zero counts, while in the ZIP model the number of non-zero counts results from both submodels. 

We ran the SAEM algorithm with 10 chains at S-step and 600 burn-in iterations followed by 100 refinement iterations. For the simulation function in the Poisson models, we truncate the counts to the maximum value of the observed counts to avoid simulating aberrant counts.

\paragraph{Results: } Table~\ref{tab:rapiModels} compares the estimates obtained with the three models in {\sf saemix}. We can compare the BIC for the Poisson (21508.6) and the ZIP (20512.9) models, which were fit on the same dataset, showing a drop of about 100 points. This improvement is born out by the better agreement between the observed proportion of zeroes, given by $p_0$ (compared to 0.15 for the Poisson model) and the observed value (0.21). The parameters estimated between the two models are very close however. Finally, the hurdle model also predicts very well the proportion of zeroes (0.21), and again the estimates of all parameters for the truncated Poisson model part, describing the non-zero counts, are very similar. Furthermore, our parameter estimates are very close to those previously reported in Tables 1 and 2 of~\citep{Atkins13} for these two models.
 
% For mixed models, the dispersion parameter can be calculated as the ratio of the sum of the squared Pearson residuals to the residual degrees of freedom (e.g., Zuur et al., 2009 p. 224). Moreover, the sum of squared Pearson residuals should approximate a Chi-squared distribution with n-p degrees of freedom (Bolker et al., 2009), and so we can also test for the ‘significance’ of overdispersion in models. 
% Similarly to categorical data, we need the value of the outcome to compute the associated likelihood. Therefore, to create the data object using {\sf saemixData}, we need to specify the response column both as a response and as a predictor. 
% How can we show overdispersion (not just lack of fit to 0's) ? $\lambda$ depends on time and many factors so not sure what to compare (variance of counts overall or need to take into account the structure of the MEM ?)
\begin{table}[!h]
\begin{center}    
    \begin{tabular}{cccc}
\hline
  & Poisson & ZIP & Hurdle \\ 
  & & (Truncated Poisson) \\
  \hline
$\alpha_0$ (-) & 1.49 & 1.57 & 1.61 \\ 
  $\beta_{male,\alpha_0}$ (-) & 0.20 & 0.19 & 0.21 \\ 
  $\alpha_1$ (mo$^{(-1)}$) & -0.04 & -0.04 & -0.02 \\ 
  $\beta_{male,\alpha_1}$ (-) & 0.02 & 0.02 & 0.01 \\ 
  $p_0$ (-)* & - & 0.08 & - \\ 
  $\omega_{\alpha_0}$ (-) & 0.96 & 0.89 & 0.85 \\ 
  $\omega_{\alpha_1}$ (-) & 0.06 & 0.06 & 0.05 \\ 
  $\rho(\alpha_0,\alpha_1)$ (-) & -0.14 & -0.09 & -0.29 \\ 
   \hline \\
   * {\it Only present in the Zero-Inflated Poisson model}
    \end{tabular}
    \caption{Parameter estimates for the Poisson, Zero-Inflated Poisson and Truncated Poisson portion of the hurdle model. The full results including uncertainty estimates as well as the logistic regression model for the zero counts are shown in the Appendix (\ref{tab:rapiModelsSE}).\label{tab:rapiModels}}
\end{center}
\end{table}
% Atkins 
% Parameter    RR     beta  95% CI
% Intercept   4.00  1.39 3.64 4.40
% Male        1.22  0.20 1.06 1.41
% Time        0.96 −0.04 0.96 0.97
% Male × Time 1.02  0.02 1.01 1.03
%             Count Sub-model
%     95% CI for RR
%             RRa      B      Lower   Upper
% Intercept   4.72     1.55    4.27    5.15
% Male        1.21     0.19    1.08    1.40
% Time        0.98    −0.02    0.98    0.99
% Male × Time 1.01     0.01    1.00    1.02
%             Logit Sub-model
%     95% CI for OR
%             ORa      B      Lower   Upper
% Intercept   4.76     1.56    3.74    6.47
% Male        0.83    −0.19    0.59    1.18
% Time        0.97    −0.03    0.95    0.99
% Male × Time 1.02     0.02    0.99    1.04

From these estimates, we can derive relative risks for the gender effects. For example, men have a RAPI score increased by 22\% ($1.22=\exp(0.20)$) at baseline compared to women. Similarly, for every month in the study, the RAPI count is expected to decrease by 4\% ($0.96=\exp(-0.04)$) for the first two models and 2\% for the Hurdle model. Note that the Hurdle model has more flexibility in modelling separately the evolution of zero and non-zero counts, as can be seen from the different estimates of the gender effect on the two sub-components (see Table~\ref{tab:rapiModelsSE} in the Appendix). Figure~\ref{fig:ZIPVPC} shows diagnostic plots for the ZIP model; VPC were obtained by simulating under the fitted model, and overlaying the empirical proportion of each count (or count category for the high counts) on the expected interval, and stratifying on gender. These plots show a good fit across all counts and indicate the model explains both the general tendency and the differences between men and women quite well. Similar diagnostic plots are given in Appendix for the Hurdle model.

\begin{figure}[!h]
    \centering
    \includegraphics[scale=0.5]{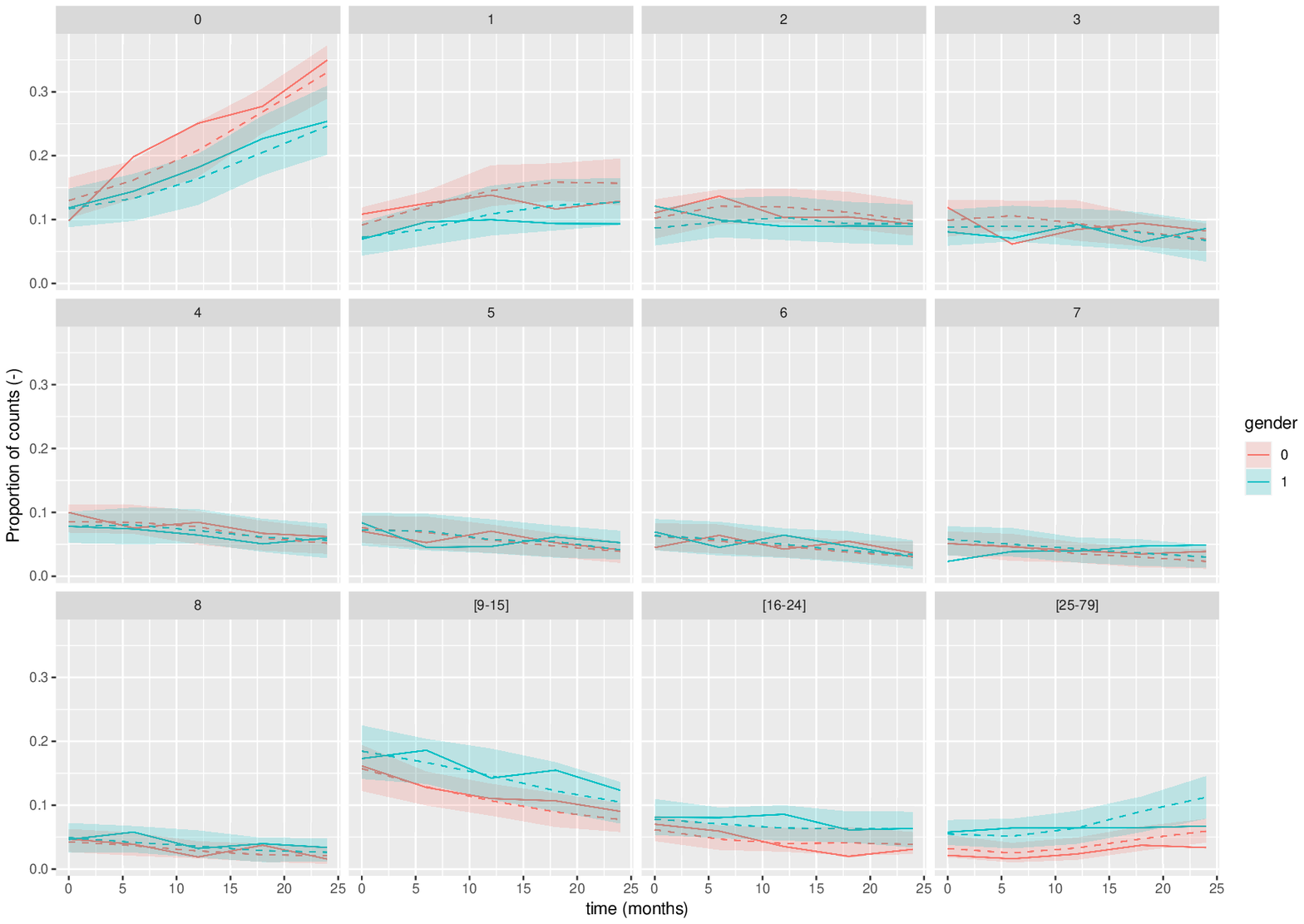}
    \caption{\label{fig:ZIPVPC} Visual Predictive Checks for the ZIP model, regrouping the high count categories and stratifying on gender.}
\end{figure}

The associated notebook on count data (see github) shows how to perform these analyses in {\sf saemix}. It also includes other possible models for overdispersion including the generalised Poisson model, which introduces an overdispersion parameter, and the negative binomial model, where a Gamma distribution is used to model the parameter $\lambda$.

\newpage
\subsection{Time-to-event data}

% Notebook on TTE. \Eco{Points of interest:}
% \begin{itemize}
% \item data exploration and model diagnostics
% \item simulation function (somewhat complicated)
% \item bootstrap to obtain SE
% \end{itemize}

\begin{figure}[!h]
    \centering
    \includegraphics[scale=0.35]{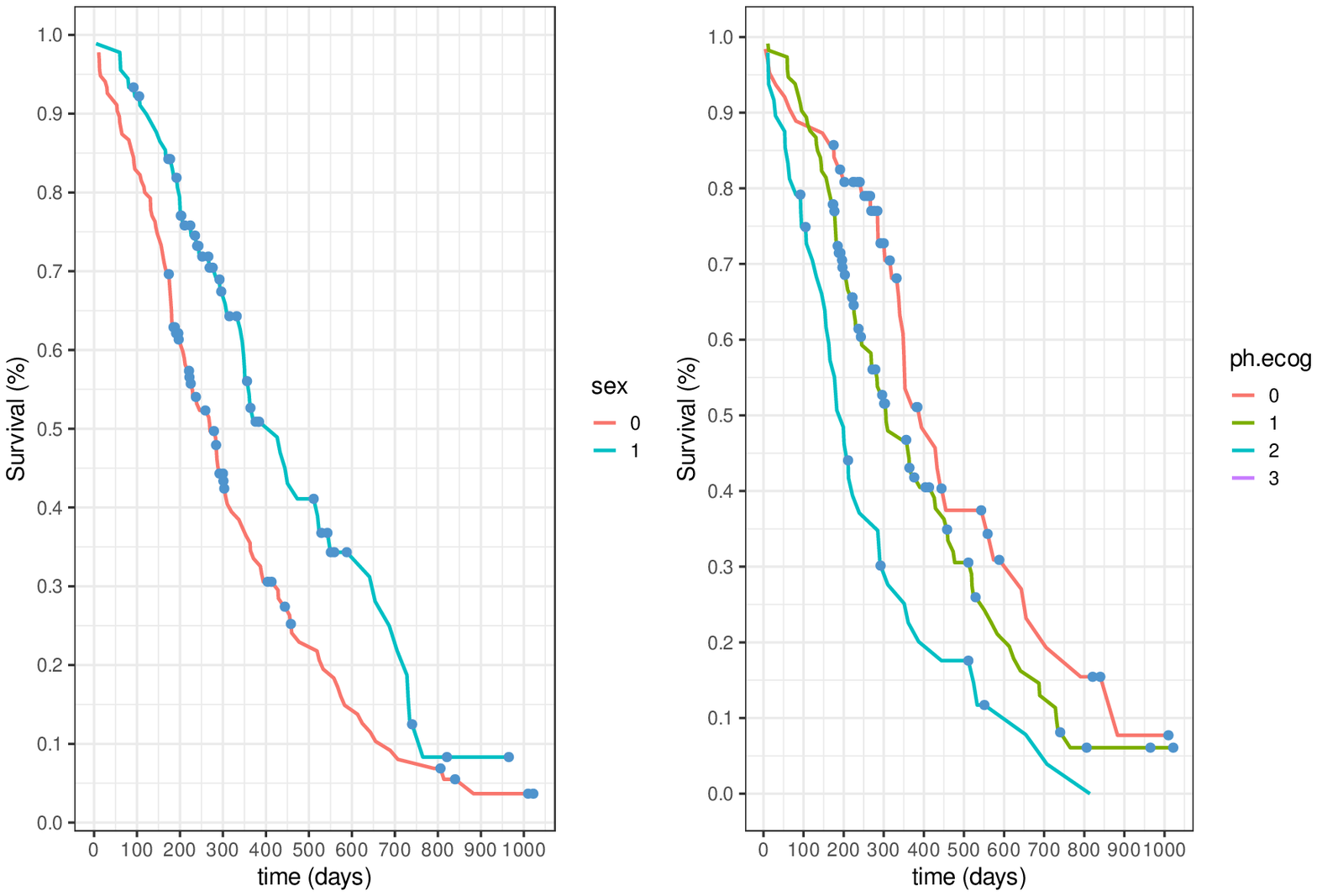}
    \caption{\label{fig:exploreTTE} Survival in the lung cancer data, stratified by gender (left) or ECOG assessment (right). The same graphs could be obtained in months instead of days by transforming the data beforehand, but since the dataset is in days we keep this unit throughout the analysis.}
\end{figure}

\paragraph{Data:} The example chosen to illustrate the analysis of time-to-event (TTE) data in {\sf saemix} is the NCCTG Lung Cancer Data, describing the survival in patients with advanced lung cancer from the North Central Cancer Treatment Group~\citep{Loprinzi94}. Covariates measured in the study include performance scores rating how well the patient can perform usual daily activities. We reformatted the {\sf cancer} dataset provided in the {\sf survival} package in R in {\sf saemix} format: patients with missing age, gender, institution or physician assessments were removed from the dataset, leaving 225 patients in the dataset out of the original 228. Status was recoded as 1 for death and 0 for a censored event, and a censoring column was added to denote whether the patient was dead or alive at the time of the last observation. A line at time=0 was added for all subjects, as the beginning of the observation period is a requisite for the algorithm. Covariates recorded for each patient included gender (0: male, 1: female), age (yr), ECOG (Eastern Cooperative Oncology Group) performance status (0-5), assessed by the physician, and Karnofsky performance status (0-100) assessed by both physician and patient. We can explore the data by plotting the survival curve, as in Figure~\ref{fig:exploreTTE} where we plot the survival stratified on gender or ECOG assessment at baseline.

\paragraph{Models:} with {\sf saemix} we can fit any parametric model for which the likelihood can be written. Here, we will fit several classical parametric survival models, the exponential, log-logistic, gamma, Weibull and Gompertz models, for which we give the equations in Table~\ref{tab:equationsTTEmodels}.
\begin{table}[!h]
\begin{center}    
    \begin{tabular}{rcccc}
  Model & Hazard function $h(t)$& Cumulative hazard $H(t)$ & Inverse survival function $S^{-1}(p)$ & Parameters\\ 
  \hline
Exponential & $\frac{1}{T_e}$ & $\frac{t}{T_e}$ & $-T_e \ln{(p)}$ & $T_e$\\
Weibull & $\frac{\gamma}{T_e} \; \left( \frac{t}{T_e} \right)^{\gamma-1}$  & $ \left( \frac{t}{T_e} \right)^\gamma$ & $T_e \; (-\ln{(p)})^{1/\gamma}$ &  $T_e, \gamma$\\
Gompertz & $\frac{\gamma}{T_e'} \; e^{\frac{t}{T_e'}}$ & $\gamma \; (e^{\frac{t}{T_e'}}-1)$ & $T_e' \; \ln{\left( 1-\frac{\ln(p)}{\gamma}\right)} $ &$T_e, \gamma$ \\
Gamma & $\frac{t^{\gamma-1} e^{-\frac{t}{T_e}}}{T_e^{\gamma} \Gamma(\gamma)}$ & $\frac{\Gamma(\gamma, t/T_e}{\Gamma(\gamma)}$ & no closed form & $T_e, \gamma$\\
Log-logistic & $ \frac{\frac{\gamma}{T_e} \left( \frac{T}{T_e} \right)^{\gamma-1}}{1+ \left( \frac{T}{T_e} \right)^{\gamma} }$ & $\ln{\left( 1+ \left(\frac{T}{T_e} \right)^{\gamma} \right)}$ & $T_e \left( e^{-\ln{(p)}}-1 \right)^{1/\gamma}$ & $T_e, \gamma$ \\
   \hline
    \end{tabular}
    \caption{Survival models tested on the lung cancer data. $T_e$ is the scale parameter and $\gamma$ controls the shape of the model. In the Gompertz model, we define the parameter $T_e'$ as $T_e'=\frac{T_e}{ln \left( 1+ \frac{ln(2)}{\gamma} \right)}$. In the Gamma model, $\Gamma(x)$ denotes the Gamma distribution and $\Gamma(x, t)$ the incomplete Gamma distribution. \label{tab:equationsTTEmodels}}
\end{center}
\end{table}
Log-normal distributions were assumed for all parameters since they should remain positive. With a single event, identifying interindividual variability is difficult, therefore we arbitrarily set it on parameter $T_e$. Model selection for the structural model was based on the BIC.

To use the automated diagnostics, a simulation function must be added to the model object. Simulating from a TTE model is slightly more complicated than for the other non Gaussian models. When the hazard function has an inverse, we can use the inverse CDF technique (or inverse transformation algorithm) to generate random samples from the TTE model. The method uses the fact that a continuous cumulative density function, $F$, is a one-to-one mapping of the domain of the cdf into the interval (0,1). Therefore, if $U$ is a uniform random variable on (0,1), then $X = F^{-1}(U)$ has the distribution $F$. 

For the single event Weibull model:
\begin{equation}
F(T)=1-e^{-\int_0^{T} h(u) du} = 1-e^{-\left( \frac{T}{\gamma} \right)^{\beta}} \sim \mathcal{U}(0,1)    
\end{equation}
and the inverse function is simply:
\begin{equation}
F^{-1}(U)=\gamma \left(-\ln(1-U) \right)^{1/\beta}   
\end{equation}
Assuming we simulate $V=1-U$ from $\mathcal{U}(0,1)$, we can obtain a sample from the Weibull parametric model as $F^{-1}(V)$. The inverse survival functions are given in Table~\ref{tab:equationsTTEmodels} for all models except the Gamma distribution for which the inverse function does not have a closed form expression. The simulations will then be used to produce Kaplan-Meier VPC~\citep{Bjornsson11}.

\paragraph{Results:} The BIC obtained with the five models are shown in table~\ref{tab:fitTTEmodels}.
\begin{table}[!h]
\begin{center}    
    \begin{tabular}{rc}
\hline
Model & BIC \\ 
  \hline
Exponential & 2303.05 \\ 
  Weibull & 2291.02 \\ 
  Gompertz & 2291.78 \\ 
  Gamma & 2378.34 \\ 
  Log-logistic & 2306.24 \\ 
   \hline
    \end{tabular}
    \caption{Comparing survival models on the lung cancer data. \label{tab:fitTTEmodels}}
\end{center}
\end{table}

\begin{figure}[!h]
    \centering
    \includegraphics[scale=0.35]{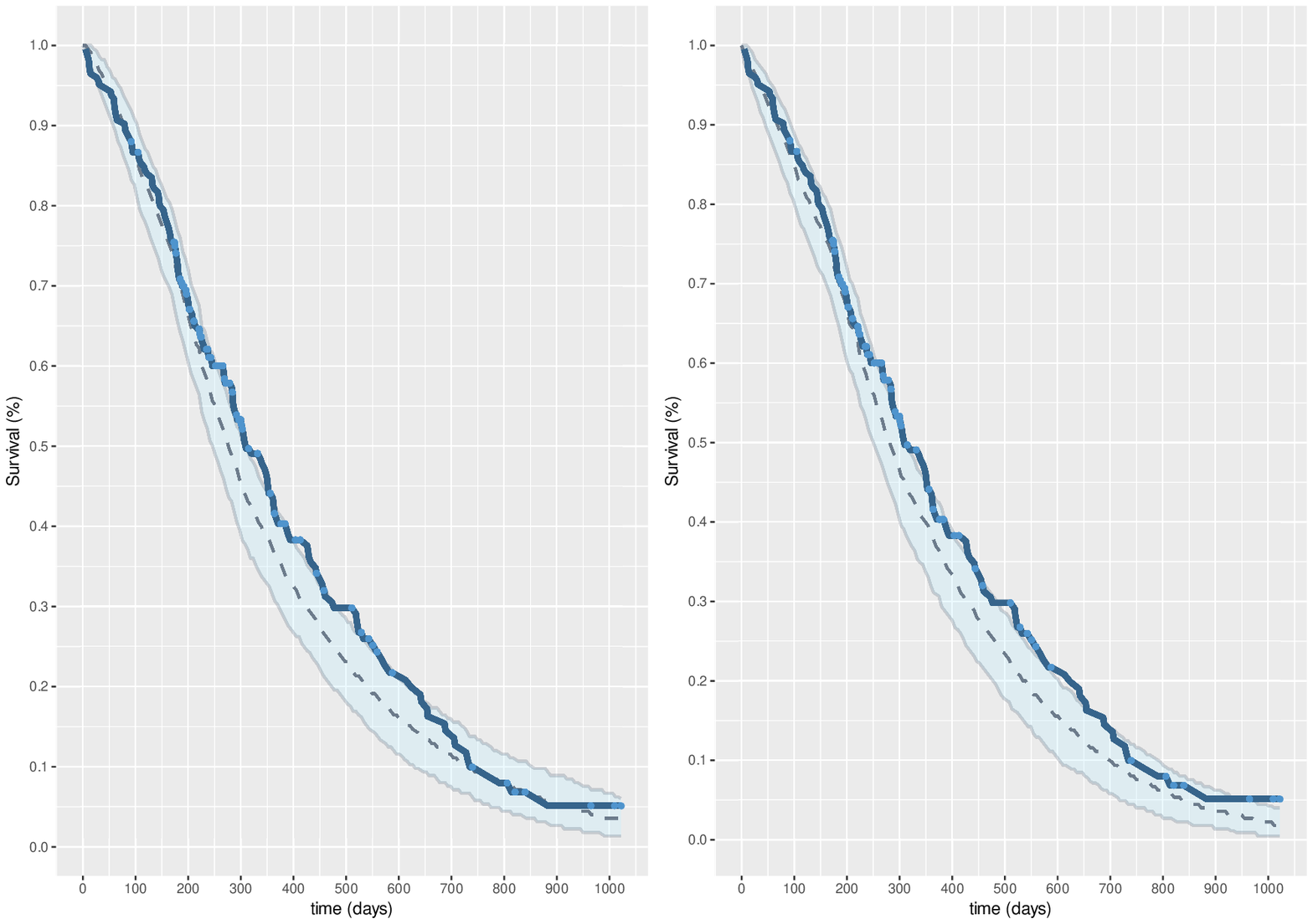}
    \caption{\label{fig:KMlung} Comparing the expected survival with the Weibull (left) and Gompertz (right) models. The shaded light blue area represents a Kaplan-Meier VPC, with the predicted survival shown as a dashed line, and the observed survival as a deep blue line.}
\end{figure}

The diagnostic plots for the Weibull and the Gompertz model, which had very similar BIC criteria, are shown in Figure~\ref{fig:KMlung}. Both fits show a slight underprediction of the survival during the second year, and the Weibull model seems to fit the end of the curve slightly better, but there are very few remaining subjects in the study at this stage. 

The covariate model was then built using the Weibull model and applying the automated algorithm~\citep{Delattre20}, regrouping ECOG scores in 3 categories (0=reference, 1 and 2-3) then in two (ECOG score $\geq2$ versus lower). The resulting model included gender and ECOG status on $T_e$. Parameter estimates for the final model are shown in Table~\ref{tab:weibull}. The model needs to include interindividual variability on one parameter to be fit as a mixed effect model, but the variability cannot be identified as each subject contributes to only one observation~\citep{Kuhn05}. For the same reason, we used the case bootstrap to compute confidence intervals as sampling random effects for the conditional bootstrap would be meaningless. The model predicts a slower progression for women, with $T_e$ increased to 619 days compared to 432 days in men (yielding median survival times of 470 days (16mo) versus 328 (11mo)), while an ECOG score of 2 or more further decreases $T_e$ in men to 264 days.

%\Eco{Parameter estimates for the IIV $\omega_{\gamma}$ ??? how to interpret them ? very small and large SE, probably non-estimable\bfseries Choose final model for table}
\begin{table}[!h]
\begin{center}    
    \begin{tabular}{lcc}
\hline
Parameter & Estimate & Bootstrap CI \\ 
  \hline
T$\_e$ (days) & 405.4 & [335.4-464.3] \\ 
  $\beta_{gender,T_e}$ (-) & 0.36 & [0.17-0.64] \\ 
  $\beta_{ECOG23,T_e}$ (-) & -0.49 & [-0.89--0.24] \\ 
  $\gamma$ (-) & 1.47 & [1.21-1.73] \\ 
  $\omega_{\gamma}$ (-) & 0.085 & [0.08-0.56] \\ 
   \hline \\
    \end{tabular}
    \caption{Parameter estimates using the Weibull model with covariates for the lung cancer data. The case bootstrap was used to provide the 95\% confidence intervals. \label{tab:weibull}}
\end{center}
\end{table}
% Case-study in Monolix (228 patients instead of 225 here)
% https://monolix.lixoft.com/case-studies/tte-model-ncctg-lung-cancer-study/
% Parameters in the same range, SE Monolix quite close to the estimates here for the fixed effects, no report of the IIV

\clearpage
\newpage
\section{Discussion}

\hskip 18pt In this paper, we present the extension of the {\sf saemix} package for {\sf R} to fit non-linear mixed effect models for non-Gaussian data using the SAEM algorithm~\citep{Kuhn05}. {\sf saemix} was  first released on CRAN in 2011, and compared to the {\sf nlme} and {\sf lme4} packages also modelling continuous outcome in~\citep{Comets17} was found to perform better than these FOCE-algorithms, with less estimation failure and better statistical performance. Models for non-Gaussian data were integrated in version 3 released in February 2022. In the present paper, we walk through several case-studies, which use datasets provided in the online help of the package.

A first type of non-Gaussian outcome is categorical data, where a limited number of categories are associated with a discrete probability distribution, with binary data as a special case with only two categories such as yes/no. In this paper we show two applications, a randomised trial of fungal toenail infection comparing two treatments and a clinical study of knee pain scores recorded in 5 modalities. The parameter estimates we found in the binary regression example were virtually identical to the estimates obtained with Monolix and similar to those reported in~\citep{Molenberghs09}, although in the latter analysis a (non-significant) treatment effect was added to the intercept. Monolix provided estimates of the SE by a stochastic method that were very close to the estimates obtained with the conditional bootstrap approach, which were also validated by the empirical SE obtained in the simulation study, suggesting that in this case the case bootstrap overestimated the SE. In~\citep{Molenberghs09} on the other hand, where numerical quadrature methods were used to fit a generalised mixed effect model, the treatment effect was not found to be significant, but a difference in an added (non-significant) treatment effect to the intercepts. The toenail data was then used as a template in a simulation study to check the performances of the algorithm. We found the estimates were unbiased in the original scenario, while bias appeared in the least identifiable parameter in a second scenario with interindividual variability. With non-Gaussian data, we use several chains to stabilise the algorithm and we checked that the results were very similar with different initial values for the parameters. Similar settings were then used in the modelling of the knee dataset with a proportional odds model. Diagnostic plots showed good fit for the two higher pain scores, with a larger reduction in pain scores for the first treatment group and an improvement in knee pain over the course of treatment. However, they also highlighted definite lack of fit for the lower pain scores. Ordinal data models present an added challenge as the models become more sensitive to initial parameter estimates. We could investigate multinomial models to allow for different time trends within each value of score, or consider different types of models such as Markov models, but these are currently out of scope in {\sf saemix}.

A second type of non-Gaussian outcome is count data, where the number of possible values may increase to infinity. The definition of the probability distribution governing the counts is very easy in {\sf saemix}, and we could fit the different models suggested in~\citep{Atkins13} with very similar results. For consistency the same interindividual variability was used in all the models, with variances on both parameters, while only one variance term was used in the hurdle model in~\citep{Atkins13}. Diagnostic plots highlighted that the overdispersion when using a Poisson model could be corrected using either the ZIP or the hurdle models. As in the categorical data example, diagnostic graphs can be used to explore the fit of the model over different scores and covariates. We also used this example to explore the performance of the fast-SAEM algorithm, previously evaluated for continuous and time-to-event data~\citep{Karimi20}, and found it worked reasonably well also (data not shown), but the performance of the underlying Laplace approximation for the conditional distributions and its sensitivity to wrong initial parameters would need to be evaluated more in depth.

%\Eco{Discussion count: Interpretation of the hurdle model value at 0 still not clear for me (expected prop of 0's doesn't match with population prediction, so could be due to the IIV here, but the parameters have a normal distribution so it should be similar...).}

Finally, we can use {\sf saemix} for single and repeated time-to-event models, using parametric models to describe the hazard function. Writing the probability in these models involves computing the survival function up to the event, which may require integrating the hazard function and incur significant computation time. The performance of {\sf saemix} for TTE was investigated in a separate work~\citep{Lavalley24}. 
%\Alex{interpretation of IIV in single-event TTE ?}

%\Eco{Discussion TTE: Comparison to Monolix: Te quite close but a bit lower (413 versus 500), $\beta$ almost exactly the same, but gamma (k in Monolix) 10 times bigger... why ? Also the automated algorithm puts variability on gamma, not Te (not tested in Monolix though). Estimates for the variability of $\omega$ are higher in Monolix (0.25 versus 0.1 on the same parameter, other parameters mostly unchanged), but not sure what this means as I still don't understand how IIV can be estimated with single events...}

% TODO: bootstrap seems to be bugged now ? (the categorical data example which used to work doesn't anymore, check)
% problem initialising categorical data models, why ?

%\Eco{TODO: fit toenail data in Monolix and compare results. Also compare stochastic SE with bootstrap SE. Discussion binary: Compared one run with Monolix and very similar estimates when using 10 chains and the same CI $\Rightarrow$ \Eco{is it worth checking that the bias is also similar across the simulations (not much bias, but a little bias on beta ?}}

%\Eco{Discussion: also applies to RTTE (see notebook + possible to add IIV); parametric models only (no Cox); performance of saemix for TTE models evaluated in a simulation study. In the current version of {\sf saemix} we do not consider interval-censored time.}

{\sf saemix} is a collaborative project, which has a dedicated development github (\url{https://github.com/saemixdevelopment/saemixextension}). We welcome new contributors, especially concerning the current limitations in the package. The structure of the code is kept sufficiently simple so that implementing new features and linking to the package remains straightforward. In particular, the algorithm has been incorporated in the {\sf mkin} package to fit models for toxicokinetic data~\citep{Rmkin}.

Because the different features in non-linear mixed effect models may become quite complex, {\sf saemix} departs from traditional {\sf R} formulas when specifying the model. This allows for more flexibility as the observation model can be defined through the structural model for continuous outcome or through the probability distribution for non-Gaussian outcomes. The data also needs to be structured through a specific function to define its hierarchical nature, much as the the {\sf groupedData} structure from the {\sf nlme} package~\citep{Rnlme}. {\sf saemix} comes with many user-friendly features tailored to each class defined in the package, to help users explore their dataset, check their models, summarise their results and diagnose their fits. Compared to the options for continuous outcomes, some features are not available yet for non Gaussian outcomes. Visual Predictive Check diagnostics can be produced, either globally or stratified on a covariate, but more complex diagnostic graphs need to be coded by the users. Simulation-based diagnostics do require the simulation function to be added in the model object, making it the responsibility of the user to define. The object produced by a call to {\sf saemix} and other functions in the package contains individual estimates and predictions on top of the parameter estimates, which can be used by {\sf R} users to produce many different graphs and statistics. In addition, scripts and examples of use are available as notebooks complementing the online help included in the package,. They show how to obtain statistical criteria, diagnostic graphs and evaluation tools that can be used for the purpose of comparing models and assessing goodness of fit.

A current issue in {\sf saemix} applied to non-Gaussian outcome is the estimation of the uncertainty. Standard errors of estimation in the package are obtained through the FO approximation to the Fisher Information Matrix computed using the conditional estimates of the individual parameters, and this approximation performs very poorly for non-Gaussian responses~\citep{Ueckert17, Riviere16}. Different methods have been proposed to estimate the exact FIM, generally using a decomposition of the expectation of the Hessian through Louis's method~\citep{Louis82}. The implementation of the stochastic computation of the FIM as in {\sf Monolix}~\citep{LavielleMonolix,Savic09} is currently a work in progress, as is the stochastic approximation of the FIM during the SAEM algorithm~\citep{Kuhn05}. An alternative computation using the equivalence between the expectation of the Hessian of the log-likelihood and minus the expectation of the product of its gradient has been recently proposed~\citep{Delattre23}, and we implemented it in a work on joint models~\citep{LavalleyPAGE23}, the code of which is available on the github. Computing the FIM is a major component of optimal design approaches, and exact FIM for non-Gaussian outcomes have been proposed by Rivière et al.~\citep{Riviere16}, who proposed a stochastic computation through a Hamiltonian Monte-Carlo approach, and Ueckert et al.~\citep{Ueckert17}), who used a numerical computation through Adaptive Gaussian Quadrature. Their code to compute the expected FIM could be harnessed to compute the observed FIM, although it may be computationally intensive to compute and sum the individual FIM. Alternatively, non-parametric methods can be used to compute the uncertainty through bootstrap or sampling importance resampling (SIR)~\citep{Dosne16}. Several versions of bootstraps adapted to non-linear mixed effect models are implemented in {\sf saemix}~\citep{Comets21} and have been used in the examples here, while code for the SIR is on the github awaiting incorporation into the package in a future version.

The absence of differential equations is one of the major limitations of {\sf saemix}. Although it is relatively easy to include differential equations within the model function in the package, the efficiency of vectorisation in {\sf R} is lost by doing so and the estimation process becomes extremely slow. Better solutions involving efficient solvers in {\sf R} or outsourcing of the computational functions to {\sf C} would be needed. An example of this approach is found in the {\sf mkin} package, which uses {\sf saemix} to fit toxicokinetic data and integrates the differential equations in {\sf C}~\citep{Rmkin}. An extension using a grid approximation is currently under development to interface with complex ODE systems defining pharmacologically-based pharmacokinetic models solved by the PKSim~\citep{PKSim} software, with a proof-of-concept application to itraconazole~\citep{Donato26}.

%This can also be an issue with time-to-event outcome, as writing the log-likelihood involves the survival function which may not always have a closed-form expression. %déjà dit
Currently, {\sf saemix} only handles single-response models. However, we have developed extensions to joint models which were evaluated on several examples involving both linear and non-linear longitudinal markers combined with time-to-event data modelled using standard parametric risk functions as well as competitive risk models~\citep{Lavalley24}. A general extension to multiple responses is under development and will be implemented in future releases of the package.

In conclusion, this extension of {\sf saemix} provides a package able to fit models describing Gaussian and non-Gaussian outcomes. The package is easy to use and performs well in a variety of models, with many detailed examples of use available both in the online help and in additional notebooks on its github. It is regularly updated, and perspectives include increased modelling capabilities, such as multiple response-models, algorithms dealing with censored data or hierarchical levels of variability, but also the implementation of user-friendly features, such as automatic transformation of categorical covariates, improved diagnostics and functions to check initial fixed effects more easily.

\section*{Acknowledgments}

The authors are indebted to Pr David Atkins who kindly allowed the RAPI data to be included in saemix. They would also like to thank Pr Marc Lavielle for creating Monolix and contributing to the first version of {\sf saemix}. {\sf saemix} is a collaborative work and we gratefully acknowledge our contributors, in particular Johannes Ranke for using {\sf saemix} in his {\sf mkin} package for toxicokinetics (\url{https://www.jrwb.de/}), Alexandra Lavalley-Morelle for implementing code extensions allowing to fit joint models (available on the github (\url{https://github.com/saemixdevelopment/saemixextension/tree/master/joint}), Lucie Fayette for comparing {\sf saemix} to {\sf Monolix} and an earlier version of {\sf Pumas} (\url{https://pumas.ai/}), and Lucas Bodelle for his current work on extending {\sf saemix} to multiple responses.

\paragraph{Statements of ethical approval:} Not applicable

\paragraph{Funding:} None

\paragraph{Competing interests:} None

% \Eco{TODO: Maud and Johannes for contributions to the saemix package if not authors}.

\section*{Code}

The {\sf saemix} package is available on CRAN (current version: 3.5), and successive versions are archived on the IAME research center github (\url{https://github.com/iame-researchCenter/saemix}), where the user guide for {\sf saemix} can be downloaded (\url{https://github.com/iame-researchCenter/saemix/blob/main/docsaem.pdf}). 

The development version of {\sf saemix} can be found on the github repository (\url{https://github.com/saemixdevelopment/saemixextension}). Notebooks and R scripts for each section of this paper have been placed in their own folder (\url{https://github.com/saemixdevelopment/saemixextension/tree/master/paperSaemix3}).

Before running the codes given in Appendix B to E, the {\sf saemix} package needs to be loaded by executing the following command:
\begin{lstlisting}[language=R]
library(saemix)
\end{lstlisting}
This should also load the required packages ({\sf npde}, {\sf ggplot2}). If not, please also load these libraries.

%\Eco{Other option: show all the raw data plots in one figure in the Appendix to illustrate the different exploratory plots available with saemix}

%\Eco{TODO: add bootstrap results for count data (en cours)}

%\Eco{TODO: finish notebook and associated R code for TTE (need to add covariate models and SE), as well as the code to generate results, figures and tables}
 
%\Eco{TODO: upload latest version of saemix + documentation + notebooks + code for paper on IAME github + clean README}

\clearpage
\newpage
%\section*{References}

\bibliographystyle{plainnat}
\bibliography{references} 

@Book{LavielleMonolix,
author        = {Marc Lavielle},
title       = {Mixed effects models for the population approach: models, tasks, methods and tools},
publisher = {Chapman \& Hall CRC Biostatistics Series},
address      = {Boca Raton, FL},
year         = 2014
}

@Book{Tutz11,
  title={Regression for Categorical Data},
  author={Tutz, Gerhard},
  isbn={9781139499583},
  series={Cambridge Series in Statistical and Probabilistic Mathematics},
  url={https://books.google.fr/books?id=hvxuqoxD00kC},
  year={2011},
  publisher={Cambridge University Press}
}

@Book{Pinheiro00,
    title = {Mixed-Effects Models in {S} and {S-PLUS}},
    author = {José C. Pinheiro and Douglas M. Bates},
    year = {2000},
    publisher = {Springer},
    address = {New York},
    doi = {10.1007/b98882},
}

@article{PKSim,
title = {{PK-Sim}: a physiologically based pharmacokinetic ‘whole-body’ model},
author = {Stefan Willmann and Jorg Lippert and Michael Sevestre and Juri Solodenko and Franco Fois and Walter Schmitt},
journal = {Biosilico},
volume = {1},
pages = {121--4},
year = 2003
}

@Manual{R,
    title = {R: A Language and Environment for Statistical Computing},
    author = {{R Development Core Team}},
    organization = {R Foundation for Statistical Computing},
    address = {Vienna, Austria},
    year = {2006},
    note = {{ISBN} 3-900051-07-0},
    url = {http://www.R-project.org},
}

@Manual{Rcatdata,
  title = {catdata: Categorical Data},
  author = {Gunther Schauberger and Gerhard Tutz},
  year = {2020},
  note = {R package version 1.2.2},
  url = {https://CRAN.R-project.org/package=catdata},
}

@Article{Rflexsurv,
    title = {{flexsurv}: A Platform for Parametric Survival Modeling in
      {R}},
    author = {Christopher Jackson},
    journal = {Journal of Statistical Software},
    year = {2016},
    volume = {70},
    number = {8},
    pages = {1--33},
    doi = {10.18637/jss.v070.i08},
  }

@Book{Rggplot2,
    author = {Hadley Wickham},
    title = {ggplot2: Elegant Graphics for Data Analysis},
    publisher = {Springer-Verlag New York},
    year = {2016},
    isbn = {978-3-319-24277-4},
    url = {https://ggplot2.tidyverse.org},
  }

@Article{Rglmm,
    title = {{MCMC} Methods for Multi-Response Generalized Linear Mixed Models: The {MCMCglmm} {R} Package},
    author = {Jarrod D. Hadfield},
    journal = {Journal of Statistical Software},
    year = {2010},
    volume = {33},
    number = {2},
    pages = {1--22},
    url = {https://www.jstatsoft.org/v33/i02/},
  }

@Manual{Rmkin,
  title = {mkin: Kinetic Evaluation of Chemical Degradation Data},
  author = {Johannes Ranke},
  year = {2020},
  note = {R package version 0.9.50.3},
  url = {https://CRAN.R-project.org/package=mkin},
}

@Manual{Rnlme,
    title = {nlme: Linear and Nonlinear Mixed Effects Models},
    author = {José Pinheiro and Douglas Bates and {R Core Team}},
    year = {2023},
    note = {R package version 3.1-162},
    url = {https://CRAN.R-project.org/package=nlme},
  }

@Misc{Rstan,
    title = {{RStan}: the {R} interface to {Stan}},
    author = {{Stan Development Team}},
    note = {R package version 2.21.8},
    year = {2023},
    url = {https://mc-stan.org/},
}

@article{Rprlogistic,
 title={Prevalence ratio estimation via logistic regression: a tool in {R}},
 volume={93},
 doi={10.1590/0001-3765202120190316},
 number={4},
 journal={Anais da Academia Brasileira de Ciencias},
 author={Leila Amorim and Raydonal Ospina},
 year={2021},
 pages={e20190316}
}

@Book{Davidian95,
     author        = {Marie Davidian and David M. Giltinan},
     title       = {Nonlinear models for repeated measurement data},
     publisher = {CRC},
     address      = {London},
     year         = 1995
}

@article{Stan,
 title={Stan: A Probabilistic Programming Language},
 volume={76},
 url={https://www.jstatsoft.org/index.php/jss/article/view/v076i01},
 doi={10.18637/jss.v076.i01},
 number={1},
 journal={Journal of Statistical Software},
 author={Carpenter, Bob and Gelman, Andrew and Hoffman, Matthew D. and Lee, Daniel and Goodrich, Ben and Betancourt, Michael and Brubaker, Marcus and Guo, Jiqiang and Li, Peter and Riddell, Allen},
 year={2017},
 pages={1–32}
}

@article{Pumas,
  title={Accelerated predictive healthcare analytics with {P}umas, a high performance pharmaceutical modeling and simulation platform},
  author={Rackauckas, Chris and Ma, Yingbo and Noack, Andreas and Dixit, Vaibhav and Mogensen, Patrick Kofod and Byrne, Simon and Maddhashiya, Shubham and Santiago Calder{\'o}n, Jos{\'e} Bayo{\'a}n and Nyberg, Joakim and Gobburu, Jogarao VS and Vijay Ivaturi},
  journal = {BioRxiv},
  url={https://doi.org/10.1101/2020.11.28.402297},
  year={2020}
}

@inproceedings{Nonmem,
title="NONMEM 7.5.1 users guides",
  author="Stuart Beal and Lewis Sheiner and Alison Boeckmann and Robert Bauer",
  year="1989-2022",
  Note="ICON plc, Gaithersburg, \url{https://nonmem.iconplc.com/nonmem751}"}

@Article{Atkins13,
author  ={David Atkins and S Baldwin and C Zheng and R Gallop and C Neighbors},
title   ={A tutorial on count regression and zero-altered count models for longitudinal substance use data},
volume  ={27},
pages   ={166–-77},
journal ={Psychology of Addictive Behaviors},
year    =2013	}

@Article{Rlme4Bates15,
  author={Bates, Douglas and M{\"a}chler, Martin and Bolker, Ben and Walker, Steve},
title   ={Fitting Linear Mixed-Effects Models Using lme4},
volume  ={67},
pages   ={1--48},
journal ={Journal of Statistical Software},
year    =2015	}

@article{Bjornsson11,
  title={Modelling of pain intensity and informative dropout in a dental pain model after naproxcinod, naproxen and placebo administration},
  author={Bj{\"o}rnsson, Marcus A and Simonsson, Ulrika SH},
  journal={British journal of Clinical Pharmacology},
  volume={71},
  number={6},
  pages={899--906},
  year={2011},
  publisher={Wiley Online Library}
}

@Article{Comets11,
author	={Emmanuelle Comets and Audrey Lavenu and Marc Lavielle},
title	={{SAEMIX}, an {R} version of the {SAEM} algorithm},
pages	={Abstr 2173},
journal	={PAGE 20},
year	=2011	}

@Article{Comets17,
author	={Emmanuelle Comets and Audrey Lavenu and Marc Lavielle},
title	={Parameter estimation in nonlinear mixed effect models using saemix, an {R} implementation of the {SAEM} algorithm},
volume	={80},
pages	={1--41},
journal	={Journal of Statistical Software},
year	=2017	}

@article{Comets08,
	title = {Computing normalised prediction distribution errors to evaluate nonlinear mixed-effect models: {The} npde add-on package for {R}},
	volume = {90},
	AAissn = {01692607},
	shorttitle = {Computing normalised prediction distribution errors to evaluate nonlinear mixed-effect models},
	number = {2},
	AAurldate = {2019-06-03},
	journal = {Computer Methods and Programs in Biomedicine},
	author = {Comets, Emmanuelle and Brendel, Karl and Mentré, France},
	AAmonth = may,
	year = {2008},
	PAGES = {154--166}
}

@Article{Comets21,
author	={Emmanuelle Comets and Christelle Rodrigues and Vincent Jullien and Moreno Ursino},
title	={Conditional non-parametric bootstrap for non-linear mixed effect models},
volume	={38},
pages	={1057--66},
journal	={Pharmaceutical Research},
year	=2021	}

@Article{CometsNpde21,
	author = {Comets, Emmanuelle and Mentr{\'e}, France},
	title = {Developing tools to evaluate non-linear mixed effect models: 20 years on the npde adventure},
	volume = {23},
	number = {4},
	journal = {The AAPS Journal},
	month = may,
	year = {2021},
	pmid = {34009502},
	pages = {75}
}

@article{debacker_toenail,
  title={Twelve weeks of continuous oral therapy for toenail onychomycosis caused by dermatophytes: a double-blind comparative trial of terbinafine 250 mg/day versus itraconazole 200 mg/day},
  author={De Backer, M and De Vroey, C and Lesaffre, Emmanuel and Scheys, I and De Keyser, P},
  journal={Journal of the American Academy of Dermatology},
  volume={38},
  number={5},
  PAGES={S57--S63},
  year={1998},
  publisher={Elsevier}
}

@Article{Delattre14,
author  ={Maud Delattre and Marc Lavielle and Marie-Anne Poursat},
title   ={A note on {BIC} in mixed effects models},
volume  ={8},
pages   ={456--475},
journal ={Electronic Journal of Statistics},
year    =2014	}

@article{Delattre20,
url = {https://doi.org/10.1515/ijb-2019-0082},
title = {An iterative algorithm for joint covariate and random effect selection in mixed effects models},
author = {Maud Delattre and Marie-Anne Poursat},
pages = {20190082},
volume = {16},
number = {2},
journal = {The International Journal of Biostatistics},
doi = {doi:10.1515/ijb-2019-0082},
year = {2020},
lastchecked = {2023-07-25}
}

@article{Delattre23,
  title={Computing an empirical {F}isher information matrix estimate in latent variable models through stochastic approximation},
  author={Delattre, Maud and Kuhn, Estelle},
  journal={Computo},
  year={2023},
  publisher={French Statistical Society}
}

@Article{Delyon99,
author  ={Bernard Delyon and Marc Lavielle and Eric Moulines},
title   ={Convergence of a stochastic approximation version of the {EM} algorithm},
volume  ={27},
pages   ={94--128},
journal ={Annals of Statistics},
year    =1999	}

@article {Dempster77,
  author={Dempster, Arthur P and Laird, Nan M and Rubin, Donald B},
     TITLE = {Maximum likelihood from incomplete data via the {E}{M} algorithm},
      NOTE = {With discussion},
   JOURNAL = {Journal of the Royal Statistical Society Series B},
    VOLUME = {39},
      YEAR = {1977},
    NUMBER = {1},
     PAGES = {1--38}
}

@Article{Desmee17,
author  ={Solène Desm{\'e}e and F Mentr{\'e'} and Christine Veyrat-Follet and Bernard S{\'e}bastien and Jérémie Guedj},
title   ={Using the SAEM algorithm for mechanistic joint models characterizing the relationship between nonlinear PSA kinetics and survival in prostate cancer patients},
volume  ={73},
pages   ={305--12},
journal ={Biometrics},
year    =2017	}

@article{Dosne16,
  title="Improving the estimation of parameter uncertainty distributions in nonlinear mixed effects models using sampling importance resampling",
  author="Anne-Gaelle Dosne  and Martin Bergstrand and Kajsa Harling and Mats O Karlsson",
  journal="Journal of Pharmacokinetics and Pharmacodynamics",
  volume="43",
  number="6",
  pages="583-596",
  year="2016",
doi="10.1007/s10928-016-9487-8"
}

@article{Donato26,
author = {Teutonico, Donato and Marchionni, David and Lavielle, Marc and Nguyen, Laurent},
title = {Integrating Population Approaches With Physiologically Based Pharmacokinetic Models: A Novel Framework for Parameter Estimation},
journal = {CPT: Pharmacometrics \& Systems Pharmacology},
volume = {15},
number = {1},
pages = {e70186},
doi = {https://doi.org/10.1002/psp4.70186},
year = {2026}
}

@Article{Guhl22,
author  ={Mélanie Guhl and Julie Bertrand and Emmanuelle Comets},
title   ={Computation of uncertainty at finite distance in non linear mixed effects models},
Note="\url{www.page-meeting.org/?abstract=10152}",
journal ={PAGE 30},
year    =2022	}

@Article{hendrickx24,
author	={Niels Hendrickx and France Mentr{\'e} and Andreas Trasch{\"u}tz and Cynthia Gagnon and Rebecca Sch{\"u}le and {ARCA Study group} and {EVIDENCE-RND consortium} and Matthis Synofzik and Emmanuelle Comets},
title	={Prediction of individual disease progression including parameter uncertainty in rare neurodegenerative diseases: the example of Autosomal-Recessive Spastic Ataxia Charlevoix Saguenay (ARSACS)},
volume	={26},
pages	={57},
journal	={The AAPS Journal},
year	=2024	}

@article{Karimi20,
  title={f-{SAEM}: A fast Stochastic Approximation of the {EM} algorithm for nonlinear mixed effects models},
  author={Belhal Karimi and Marc Lavielle and Eric Moulines},
  journal = {Computational Statistics \& Data Analysis},
  volume = {141},
  pages={123-38}, 
  year={2020}
}

@Article{Kuhn05,
author  ={Estelle Kuhn and Marc Lavielle},
title   ={Maximum likelihood estimation in nonlinear mixed effects models},
volume  ={49},
pages   ={1020--1038},
journal ={Computational Statistics and Data Analysis},
year    =2005	}

@Article{LavalleyPAGE23,
author  ={Alexandra Lavalley-Morelle and France Mentr\'e and Emmanuelle Comets and Jimmy Mullaert},
title   ={Joint modeling of biomarkers dynamics and survival with competing risks to predict the prognostic of patients hospitalized with severe infectious diseases},
Note="\url{www.page-meeting.org/?abstract=10394}",
journal ={PAGE 31},
year    =2023	}

@Article{Lavalley24,
author  ={Alexandra Lavalley-Morelle and France Mentr\'e and Emmanuelle Comets and Jimmy Mullaert},
title   ={Extending the code in the open-source saemix package to fit joint models},
journal ={Computer Methods and Programs in Biomedicine},
volume	={247},
pages	={108095},
year    =2024	}

@article{lesaffre2001effect,
  title={On the effect of the number of quadrature points in a logistic random effects model: an example},
  author={Lesaffre, Emmanuel and Spiessens, Bart},
  journal={Journal of the Royal Statistical Society: Series C (Applied Statistics)},
  volume={50},
  number={3},
  PAGES={325--335},
  year={2001},
  publisher={Wiley Online Library}
}

@article{lin2011goodness,
  title={A goodness-of-fit test for logistic-normal models using nonparametric smoothing method},
  author={Lin, Kuo-Chin and Chen, Yi-Ju},
  journal={Journal of Statistical Planning and Inference},
  volume={141},
  number={2},
  PAGES={1069--1076},
  year={2011},
  publisher={Elsevier}
}

@article{lindstrom90,
	author = {Mary J. Lindstrom and Douglas M. Bates},
	issn = {0006341X, 15410420},
	journal = {Biometrics},
	number = {3},
	pages = {673–687},
	publisher = {[Wiley, International Biometric Society]},
	title = {Nonlinear Mixed Effects Models for Repeated Measures Data},
	url = {http://www.jstor.org/stable/2532087},
	volume = {46},
	year = {1990}
}

@Article{Loprinzi94,
  author={Loprinzi, Charles Lawrence and Laurie, John A and Wieand, H Sam and Krook, James E and Novotny, Paul J and Kugler, John W and Bartel, Joan and Law, Marlys and Bateman, Marilyn and Klatt, Nancy E},
title   ={Prospective evaluation of prognostic variables from patient-completed questionnaires. North Central Cancer Treatment Group},
volume  ={12},
pages   ={601--7},
journal ={Journal of Clinical Oncology : Official Journal of the American Society of Clinical Oncology},
year    =1994	}

@article{Louis82,
	title = {Finding the {Observed} {Information} {Matrix} {When} {Using} the {EM} {Algorithm}},
	volume = {44},
	issn = {2517-6161},
	number = {2},
	urldate = {2023-04-26},
	journal = {Journal of the Royal Statistical Society: Series B (Methodological)},
	author = {Louis, Thomas A.},
	year = {1982},
	pages = {226--233}
}

@Article{Molenberghs09,
author  ={Geert Molenberghs and Geert Verbeke},
title   ={Longitudinal and incomplete clinical studies},
volume  ={22},
pages   ={1--32},
journal ={Journal of the Japanese Society of Computational Statistics},
year    =2009	}

@Article{Neighbors10a,
author  ={Clayton Neighbors and Melissa A. Lewis and David C. Atkins and Megan M. Jensen and Theresa Walter and
Nicole Fossos and Christine M. Lee and Mary E. Larimer},
title   ={Efficacy of web-based personalized normative feedback: A two-year randomized controlled trial},
volume  ={78},
pages   ={898--911},
journal ={Journal of Consulting and Clinical Psychology},
year    =2010	}

@Article{Neighbors10b,
  title={Cost-effectiveness of a motivational lntervention for alcohol-involved youth in a hospital emergency department},
  author={Neighbors, Charles J and Barnett, Nancy P and Rohsenow, Damaris J and Colby, Suzanne M and Monti, Peter M},
  journal={Journal of Studies on Alcohol and Drugs},
  volume={71},
  number={3},
  pages={384--394},
  year={2010},
  publisher={Rutgers University}
}

@article{Nguyen17,
  title={Model evaluation of continuous data pharmacometric models: metrics and graphics},
  author={Nguyen, THT and Mouksassi, Samer and Holford, Nicholas and Al-Huniti, N and Freedman, Imanuel and Hooker, Andrew C and John, J and Karlsson, Mats O and Mould, Diane R and P{\'e}rez Ruixo, JJ and others},
  journal={CPT: pharmacometrics \& systems pharmacology},
	volume = {6},
	issn = {2163-8306, 2163-8306},
	shorttitle = {Model Evaluation of Continuous Data Pharmacometric Models},
	doi = {10.1002/psp4.12161},
	language = {en},
	number = {2},
	urldate = {2019-06-03},
	year = {2017},
	pages = {87--109}
}

@article{Riviere16,
author = {Riviere, Marie-Karelle and Ueckert, Sebastian and Mentré, France},
year = {2016},
month = {05},
pages = {kxw020},
title = {An {MCMC} method for the evaluation of the {F}isher information matrix for non-linear mixed effect models},
volume = {17},
journal = {Biostatistics}
}

@article{Savic09,
	title = {Performance in population models for count data, part {II}: {A} new {SAEM} algorithm},
	volume = {36},
	journal = {Journal of Pharmacokinetics and Pharmacodynamics},
	author = {Radojka M Savic and Marc Lavielle},
	year = {2009},
	pages = {367--79}
}

@article{Savic10,
	title = {Implementation and evaluation of the {SAEM} algorithm for longitudinal ordered categorical data with an illustration in pharmacokinetics–pharmacodynamics},
	volume = {13},
	issn = {1550-7416},
	doi = {10.1208/s12248-010-9238-5},
    number = {1},
	urldate = {2019-07-16},
	journal = {The AAPS Journal},
	author = {Savic, Radojka M. and Mentr{\'e}, France and Lavielle, Marc},
	month = nov,
	year = {2010},
	pmid = {21063925},
	pmcid = {PMC3032088},
	pages = {44--53}
	
}

@article{Ueckert17,
	title = {A new method for evaluation of the {Fisher} information matrix for discrete mixed effect models using {Monte} {Carlo} sampling and adaptive {Gaussian} quadrature},
	volume = {111},
	journal = {Computational Statistics \& Data Analysis},
	author = {Ueckert, Sebastian and Mentre, France},
	year = {2017},
	pages = {203--219}
}

@Article{White89,
  author={White, Helene R and Labouvie, Erich W},
title   ={Towards the assessment of adolescent problem drinking},
volume  ={50},
pages   ={30--7},
journal ={Journal of Studies on Alcohol},
year    =1989	}

@article {Wu83,
    AUTHOR = {Wu, C.-F. Jeff},
     TITLE = {On the convergence properties of the {E}{M} algorithm},
  JOURNAL = {The Annals of Statistics},
    VOLUME = {11},
      YEAR = {1983},
    NUMBER = {1},
     PAGES = {95--103}
}

\clearpage
\newpage
\section*{Appendix A - The SAEM algorithm}

We aim to maximize the log-likelihood of $\theta$ in equation~\ref{eq:likeNLMEM}
This expression has no closed-form in NLMEM because it requires an integration over the unknown random effects. The complete log-likelihood associated to the complete data $(y,\psi_i)$ however can be written explicitly:
$$L(\theta | y, \psi) = \sum_{i=1}^N \log \left( p(y_i, \psi_i | \theta) \right) = \sum_{i=1}^N \log \left( p(\psi_i |y_i, \theta) \right) + \log \left( p(y_i | \theta) \right)$$
Thus, we have:
$$L(\theta | y) = L(\theta | y, \psi) - \sum_{i=1}^N \log \left( p(y_i | \theta) \right) $$

As $L(\theta | y) $ does not depend on $\psi$, the expectation on $\psi$ conditionally to $\theta^{(k)}$ of the previous equality writes:

$$L(\theta | y) = \mathbb{E}_\psi \left[ L(\theta | y, \psi) | \theta^{(k)} \right] - \mathbb{E}_\psi \left[ \sum_{i=1}^N \log \left( p(y_i | \theta) \right)| \theta^{(k)} \right]  $$

Then, it can be shown that taking $\theta^{(k+1)} = \underset{ \theta }{\text{arg max} }\ \mathbb{E}_\psi \left[ L(\theta | y, \psi) | \theta^{(k)} \right] $ allows $L(\theta^{(k)} | y)$ to converge to a local maximum.
\\

Moreover, 
\begin{align*}
\mathbb{E}_\psi \left[ L(\theta | y, \psi) | \theta^{(k)} \right] & = \int L(\theta  | y, \psi_i) p(\psi_i | \theta^{(k)})  d\psi_i \\
 & = \int \sum_{i=1}^N \log \left( p(y_i, \psi_i | \theta) \right)  p(\psi_i | \theta^{(k)}) d\psi_i \\
 & = \mathbb{E}_\psi \left[ \log \left( p(y, \psi |  \theta \right) | y, \theta^{(k)} \right]
\end{align*} 

% $$ \mathbb{E}_\psi \left[ L(\theta | y, \psi) | \theta^{(k)} \right] = \int L(\theta  | y, \psi_i) p(\psi_i | \theta^{(k)})  d\psi_i = \int \sum_{i=1}^N \log \left( p(y_i, \psi_i | \theta) \right)  p(\psi_i | \theta^{(k)}) d\psi_i  = \mathbb{E}_\psi \left[ \log \left( p(y, \psi |  \theta \right) | y, \theta^{(k)} \right] $$

Thus, at iteration $k$ of EM, the conditional expectation of the complete log-likelihood $$Q_k(\theta) = \mathbb{E}\left[ \log \left( p(y, \psi; \theta ) \right)| y, \theta^{(k-1)}  \right]$$ is first computed during E-step, then the next $\theta^{(k)}$ is computed in M-step in order to maximize $Q_k(\theta)$. 

%\section*{Appendix B - Bootstrap algorithms in {\sf saemix}}
%\Eco{remove and refer to to the paper}

All the codes in the Appendix need the {\sf saemix} package to be loaded. Please execute the following commands before running the codes given in Appendix B to E.
\begin{lstlisting}[language=R]
library(saemix)
\end{lstlisting}
This should also load the required packages ({\sf npde}, {\sf ggplot2}). If not, please also load these libraries.

\newpage
\section*{Appendix B - Analysis of binary data}

\subsection*{saemix code for the toenail example}

\begin{lstlisting}[language=R]
data(toenail.saemix)
saemix.data<-saemixData(name.data=toenail.saemix,name.group=c("id"),
  name.predictors=c("time","y"), name.response="y",name.covariates=c("treatment"), 
  units=list(x="d",y="-"), verbose=FALSE)
# model and simulation functions
binary.model<-function(psi,id,xidep) {
  tim<-xidep[,1]
  y<-xidep[,2]
  inter<-psi[id,1]
  slope<-psi[id,2]
  logit<-inter+slope*tim
  pevent<-exp(logit)/(1+exp(logit))
  logpdf<-rep(0,length(tim))
  P.obs = (y==0)*(1-pevent)+(y==1)*pevent
  logpdf <- log(P.obs)
  return(logpdf)
}
simulBinary<-function(psi,id,xidep) {
  tim<-xidep[,1]
  y<-xidep[,2]
  inter<-psi[id,1]
  slope<-psi[id,2]
  logit<-inter+slope*tim
  pevent<-1/(1+exp(-logit))
  ysim<-rbinom(length(tim),size=1, prob=pevent)
  return(ysim)
}
# saemix model
saemix.model<-saemixModel(model=binary.model,description="Binary model",
  simulate.function=simulBinary, modeltype="likelihood",
  psi0=matrix(c(-0.5,-.15,0,0),ncol=2,byrow=TRUE,
  dimnames=list(NULL,c("theta1","theta2"))),
  transform.par=c(0,0), covariate.model=c(0,1),
  covariance.model=matrix(c(1,0,0,0),ncol=2), 
  omega.init=diag(c(0.5,0.3)), verbose=FALSE)

# saemix fit
saemix.options<-list(seed=1234567,save=FALSE,save.graphs=FALSE, 
  displayProgress=FALSE, nb.chains=10, fim=FALSE, print=FALSE)
binary.fit<-saemix(saemix.model,saemix.data,saemix.options)
summary(binary.fit)

# Diagnostics
nsim<-1000
binary.fit <- simulateDiscreteSaemix(binary.fit, nsim=nsim)
discreteVPC(binary.fit, outcome="binary", which.cov="treatment")
\end{lstlisting}

\subsection*{Additional results for the simulation with binary outcome}

\subsubsection*{Bias and RRMSE}
Table~\ref{tab:binSimulREE} shows the relative bias (\%) and RRMSE (\%, in brackets) for all the parameters estimated in the two scenarios. For each scenario, we tested 3 initial conditions, but the results were in close agreement regardless of the setting.
\begin{table}[!h]
    \centering
\begin{tabular}{rcccccc}
\hline
{\bf Parameter} & \multicolumn{3}{c}{\bf Original model} & \multicolumn{3}{c}{\bf IIV on both parameters} \\ 
 & {\bf True} & {\bf Pop} & {\bf Far} & {\bf True} & {\bf Pop} & {\bf Far} \\
  \hline
$\theta_1$ &  2.06 (19) &  2.07 (19) &  2.07 (19) & -1.76 (11) & -0.60 (11) & -1.65 (11) \\ 
  $\theta_2$ & -0.92 (12) & -0.91 (12) & -0.92 (12) &  7.57 (21) &  7.49 (20) &  9.22 (21) \\ 
  $\beta$ &  3.94 (47) &  3.91 (47) &  3.96 (47) & 11.57 (56) & 12.38 (56) & 12.50 (57) \\ 
  $\omega_1$ &  0.85 (10) &  0.84 (10) &  0.85 (10) & -0.89 (20) &  0.39 (20) & -0.19 (20) \\ 
  $\omega_2$ &  &  &  &  3.29 (29) &  5.19 (25) &  6.47 (26) \\ 
   \hline
\end{tabular}
\caption{Bias (\%) (RRMSE, \%) for the parameters in the two scenarios, when starting from different sets of initial values (True: parameters used for the simulation of the data; Pop: parameters initialised as in the real dataset ($\theta_1=-0.5, \theta_2=-0.19, \beta=0, \omega_1=1, \omega_2=1$); Far: fixed effects initialised to 0, and variabilities to $\omega_1=2, \omega_2=0.7$).} \label{tab:binSimulREE}
\end{table}

\subsubsection*{Violin plots}

\begin{figure}[!h]
    \centering
    \includegraphics[scale=0.4]{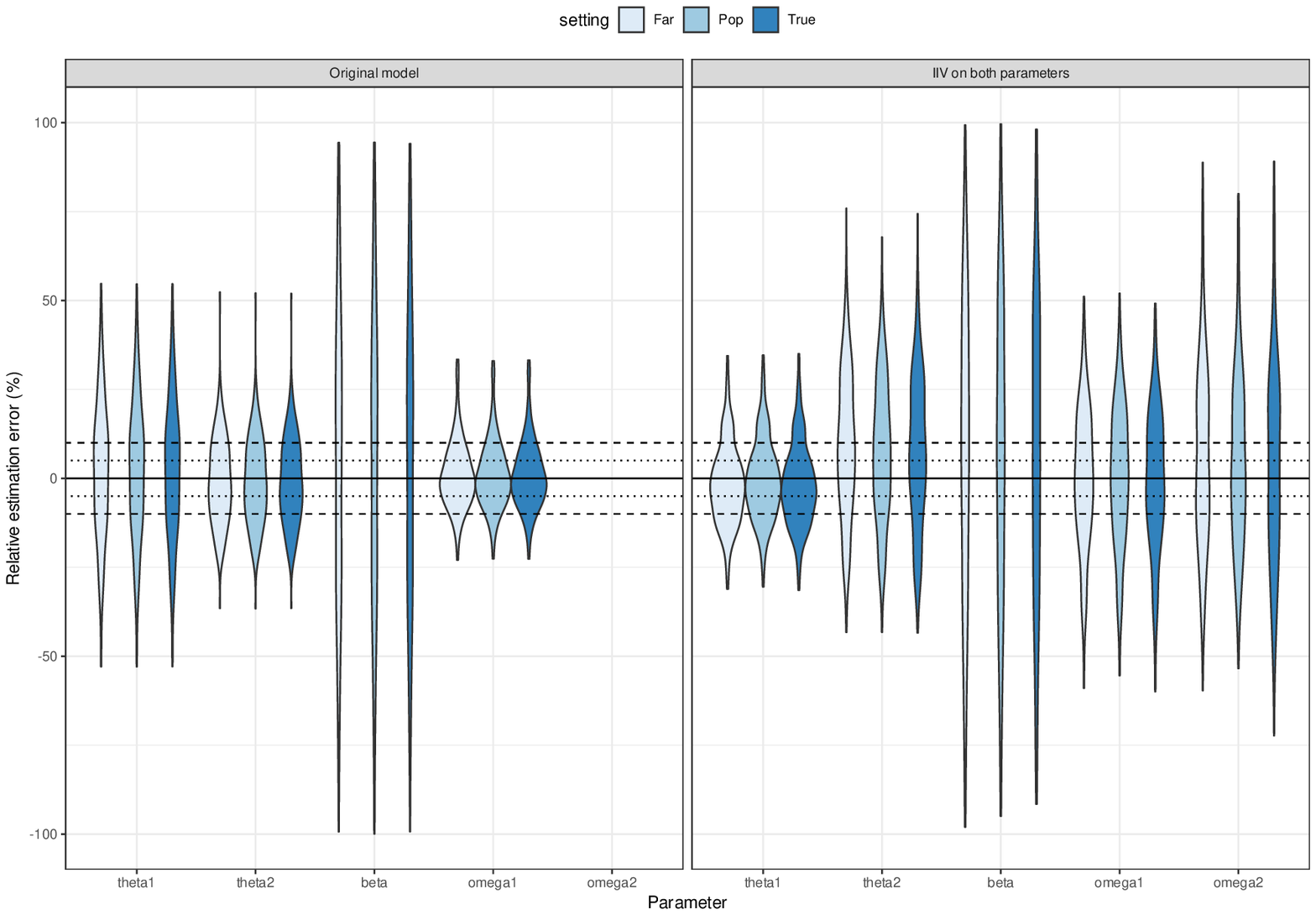}
    \caption{\label{fig:binSimulREESupp} Relative estimation errors for the parameters in the original design (left) and the design with variability on both parameters (right) with different initial conditions. Dashed lines delineate absolute relative biases within 10\% and dotted lines denote biases within 5\%.  The graph was trimmed to +/-100, omitting 78 values for $\beta$ in total over the 3 analyses.}
\end{figure}

Figure~\ref{fig:binSimulREESupp} shows the violin plots of Figure~\ref{fig:binSimulREE} for the three different settings. As in the previous table, this plot shows that the initial conditions don't have a major impact on the parameter estimates.

\clearpage
\newpage

\section*{Appendix C -  Categorical data}

\subsection*{saemix code for the analysis of knee pain scores}

\begin{lstlisting}[language=R]
data(knee.saemix)

ordknee.data<-saemixData(name.data=knee.saemix,name.group=c("id"),
   name.predictors=c("y", "time"), name.X=c("time"),
   name.covariates = c("Age","Sex","treatment","Age2"),
   units=list(x="d",y="", covariates=c("yr","-","-","yr2")), verbose=FALSE)

plotDiscreteData(ordknee.data, outcome="categorical", which.cov="treatment")
plotDiscreteData(ordknee.data, outcome="categorical", which.cov="Sex")

# Model for ordinal responses
ordinal.model<-function(psi,id,xidep) {
  y<-xidep[,1]
  time<-xidep[,2]
  alp1<-psi[id,1]
  alp2<-psi[id,2]
  alp3<-psi[id,3]
  alp4<-psi[id,4]
  beta<-psi[id,5]
  
  logit1<-alp1 + beta*time
  logit2<-logit1+alp2
  logit3<-logit2+alp3
  logit4<-logit3+alp4
  pge1<-exp(logit1)/(1+exp(logit1))
  pge2<-exp(logit2)/(1+exp(logit2))
  pge3<-exp(logit3)/(1+exp(logit3))
  pge4<-exp(logit4)/(1+exp(logit4))
  pobs = (y==1)*pge1+(y==2)*(pge2 - pge1)+(y==3)*(pge3 - pge2)+
    (y==4)*(pge4 - pge3)+(y==5)*(1 - pge4)
  logpdf <- log(pobs)
  
  return(logpdf)
}
# simulate function
simulateOrdinal<-function(psi,id,xidep) {
  y<-xidep[,1]
  time<-xidep[,2]
  alp1<-psi[id,1]
  alp2<-psi[id,2]
  alp3<-psi[id,3]
  alp4<-psi[id,4]
  beta<-psi[id,5]
  
  logit1<-alp1 + beta*time
  logit2<-logit1+alp2
  logit3<-logit2+alp3
  logit4<-logit3+alp4
  pge1<-exp(logit1)/(1+exp(logit1))
  pge2<-exp(logit2)/(1+exp(logit2))
  pge3<-exp(logit3)/(1+exp(logit3))
  pge4<-exp(logit4)/(1+exp(logit4))
  x<-runif(length(time))
  ysim<-1+as.integer(x>pge1)+as.integer(x>pge2)+as.integer(x>pge3)+as.integer(x>pge4)
  return(ysim)
}
saemix.options<-list(seed=632545,save=FALSE,save.graphs=FALSE, fim=FALSE, 
   nb.chains=10, nbiter.saemix=c(600,100), print=FALSE)

# No covariates
saemix.model<-saemixModel(model=ordinal.model,description="Ordinal categorical model",
  modeltype="likelihood", simulate.function=simulateOrdinal, 
  psi0=matrix(c(0,0.2, 0.6, 3, 0.2),ncol=5, byrow=TRUE,  dimnames=list(NULL,
  c("alp1","alp2","alp3","alp4","beta"))), transform.par=c(0,1,1,1,1),
  omega.init=diag(c(100, 1, 1, 1, 1)), 
  covariance.model = diag(c(1,0,0,0,1)), verbose=FALSE)

ord.fit<-saemix(saemix.model,ordknee.data,saemix.options)
summary(ord.fit)

# Covariate model
# Do not run, Rstudio fails (ran in a script as "R CMD BATCH paper_kneeCovModel.R paper_kneeCovModel.out")
#if(runCovKnee) cov.ordfit <- step.saemix(ord.fit, trace=TRUE, direction='both')

# Resulting model
## IIV: all alphas, none on beta :-/
## Covariates:   alp1(Age2)alp2(treatment)beta(treatment)
covariate.model <- matrix(data=0, nrow=4, ncol=5)
covariate.model[3,2]<-covariate.model[3,5]<-covariate.model[4,1]<-1
ordmodel.cov<-saemixModel(model=ordinal.model,description="Ordinal categorical model",modeltype="likelihood",
  simulate.function=simulateOrdinal, psi0=matrix(c(0,0.2, 0.6, 3, 0.2),ncol=5, byrow=TRUE, 
  dimnames=list(NULL,c("alp1","alp2","alp3","alp4","beta"))), transform.par=c(0,1,1,1,1),
  omega.init=diag(c(100, 1, 1, 1, 1)), covariate.model=covariate.model, covariance.model = diag(c(1,1,1,1,0)), verbose=FALSE)
saemix.options<-list(seed=632545,save=FALSE,save.graphs=FALSE, fim=FALSE, nb.chains=10, nbiter.saemix=c(600,100), print=FALSE)
ord.fit.cov<-saemix(ordmodel.cov,ordknee.data,saemix.options)
summary(ord.fit.cov)

# Compare the base and covariate model 
compare.saemix(ord.fit, ord.fit.cov)

# Diagnostics
nsim<-100
yfit<-ord.fit.cov
yfit<-simulateDiscreteSaemix(yfit, nsim=nsim)
discreteVPC(yfit, outcome="categorical")
plot1 <- discreteVPC(yfit, outcome='categorical',covsplit=TRUE, which.cov="treatment")
plot2 <-  discreteVPC(yfit, outcome='categorical',covsplit=TRUE, which.cov="Sex")
\end{lstlisting}

\subsection*{Additional results for the knee data}

\begin{figure}[!h]
    \centering
    \includegraphics[scale=0.4]{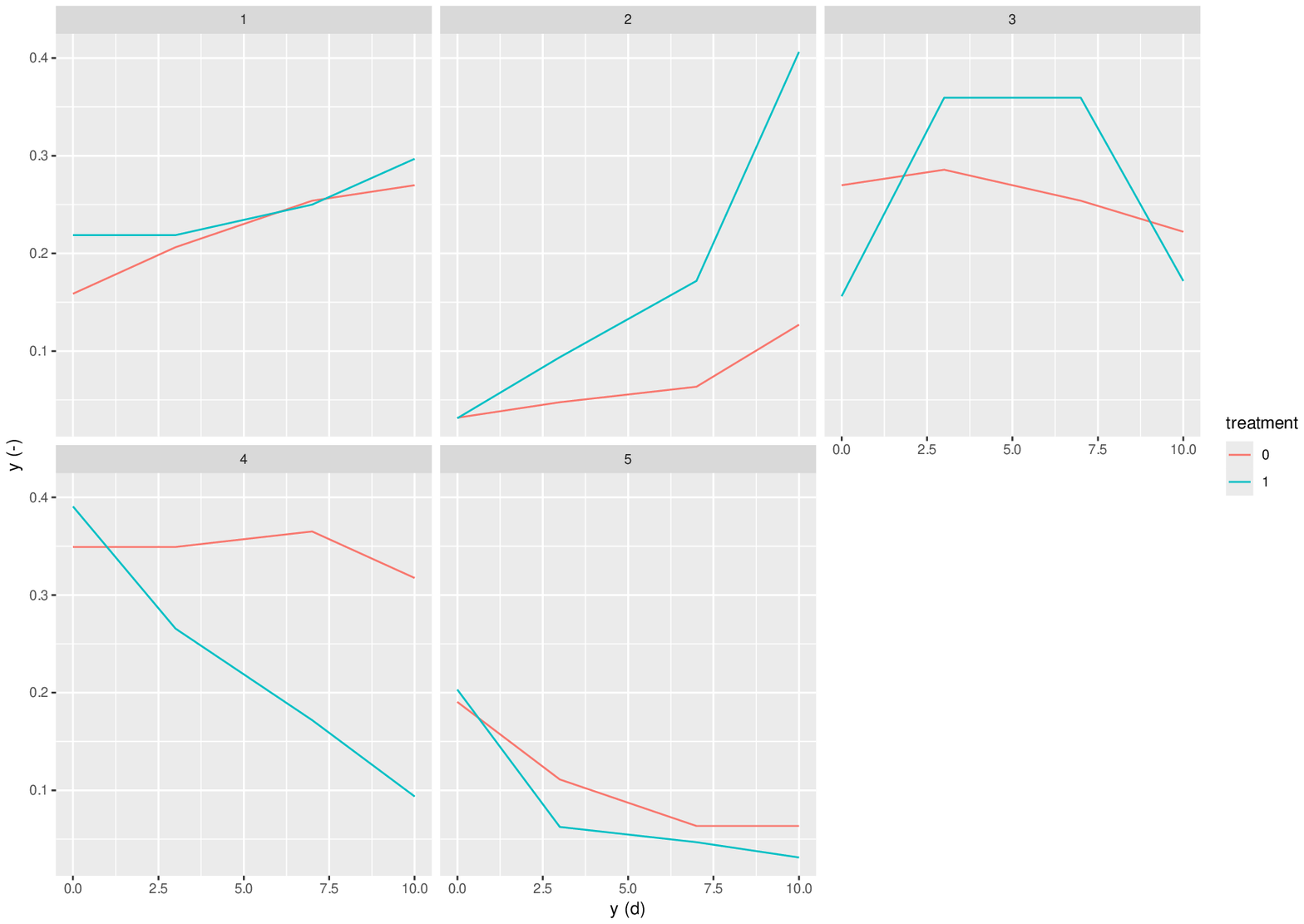}
    \caption{\label{fig:dataKnee} Evolution of the observed proportion of each score over time, stratified by treatment group.}
\end{figure}

Figure~\ref{fig:dataKnee} shows the proportion of pain scores over time, stratified by treatment group, for the knee data.

\begin{figure}[!h]
    \centering
    \includegraphics[scale=0.4]{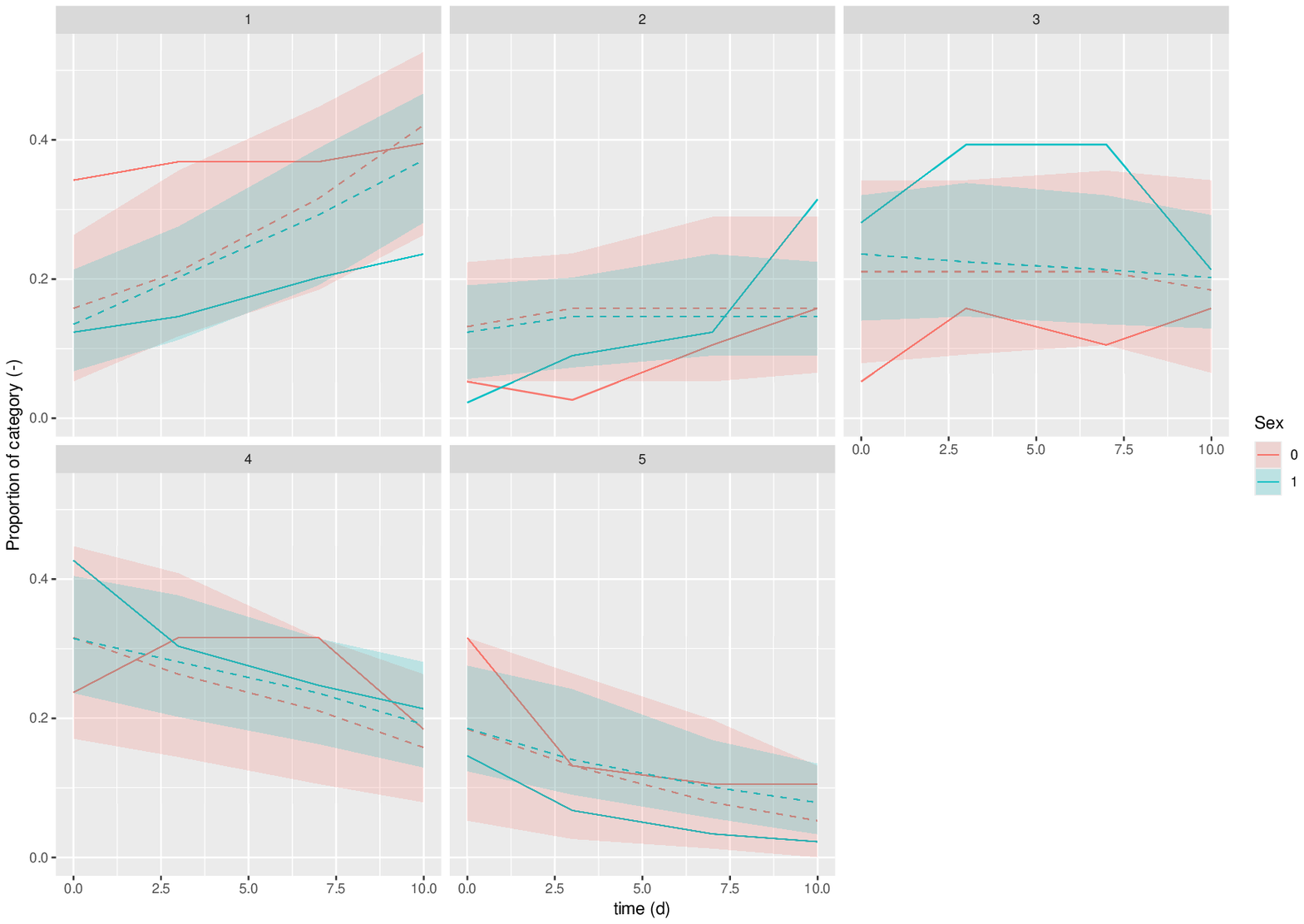}
    \caption{\label{fig:kneeVPCSex} Visual Predictive Check for each value of the score in the proportional odds model, stratified by gender.}
\end{figure}

\begin{figure}[!h]
    \centering
    \includegraphics[scale=0.4]{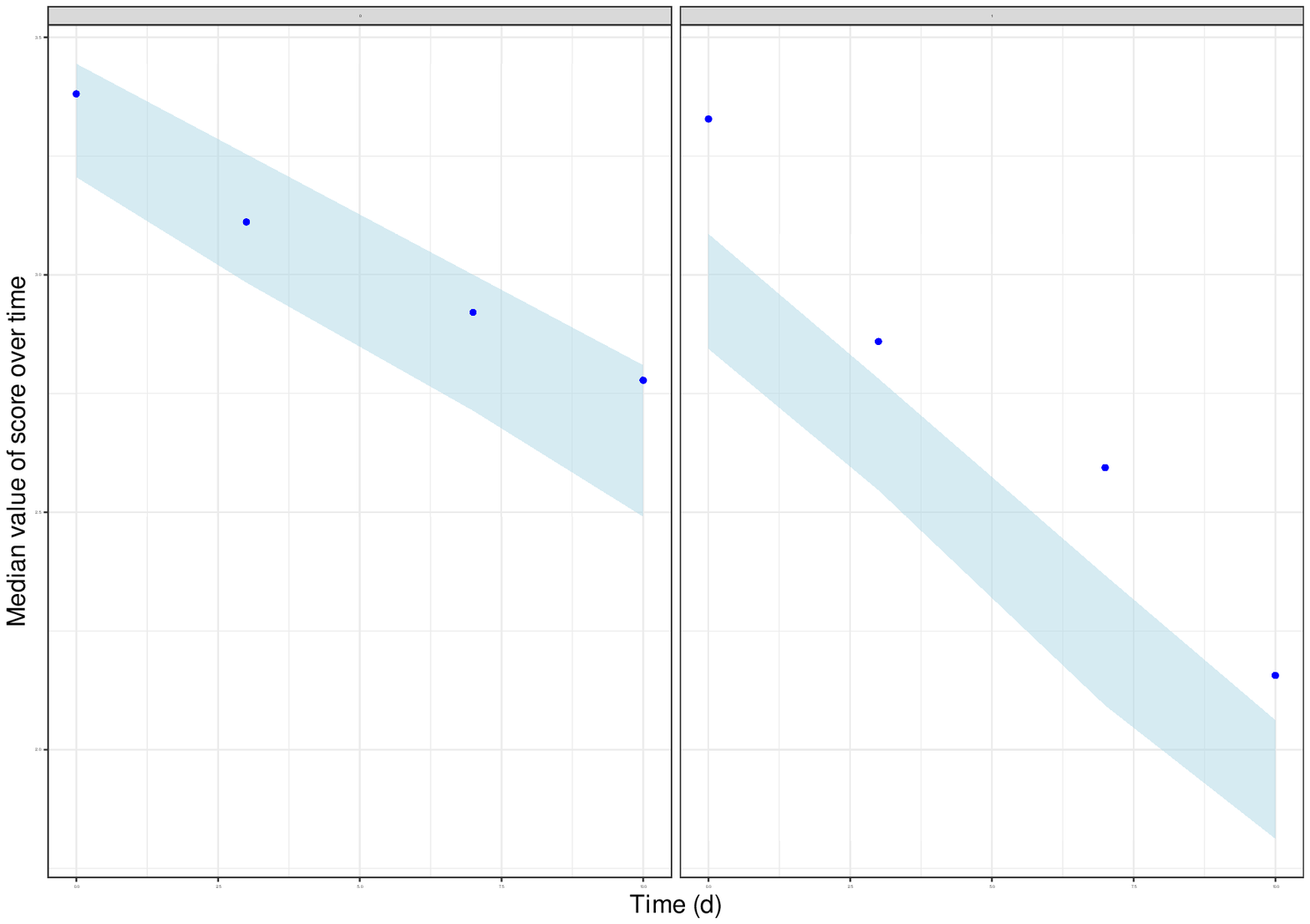}
    \caption{\label{fig:kneeVPCmedian} Visual Predictive Check for the mean score at each time point.}
\end{figure}

Additional diagnostics for model evaluation are given in Figure~\ref{fig:kneeVPCSex}, with the pain scores stratified by gender, and~\ref{fig:kneeVPCmedian}, showing the mean pain score for each treatment along with its prediction interval.

\section*{Appendix D - Count data models}

\subsection*{saemix code}

\begin{lstlisting}[language=R]
data(rapi.saemix)
rapi.saemix$gender <- ifelse(rapi.saemix$gender=="Men",1,0) # Female=reference class as in Atkins

saemix.data<-saemixData(name.data=rapi.saemix, name.group=c("id"),
  name.predictors=c("time","rapi"),name.response=c("rapi"),
  name.covariates=c("gender"),units=list(x="months",y="",covariates=c("")), verbose=FALSE)
plotDiscreteData(saemix.data, outcome="count", which.cov="gender", breaks=c(0:9, 16, 25,80))

## Zero-inflated Poisson model
count.poissonzip<-function(psi,id,xidep) {
  time<-xidep[,1]
  y<-xidep[,2]
  intercept<-psi[id,1]
  slope<-psi[id,2]
  p0<-psi[id,3] # Probability of zero's
  lambda<- exp(intercept + slope*time)
  logp <- log(1-p0) -lambda + y*log(lambda) - log(factorial(y)) # Poisson
  logp0 <- log(p0+(1-p0)*exp(-lambda)) # Zeroes
  logp[y==0]<-logp0[y==0]
  return(logp)
}
# Simulation function
countsimulate.poissonzip<-function(psi, id, xidep) {
  time<-xidep[,1]
  y<-xidep[,2]
  ymax<-max(y)
  intercept<-psi[id,1]
  slope<-psi[id,2]
  p0<-psi[id,3] # Probability of zero's
  lambda<- exp(intercept + slope*time)
  prob0<-rbinom(length(time), size=1, prob=p0)
  y<-rpois(length(time), lambda=lambda)
  y[prob0==1]<-0
  y[y>ymax]<-ymax+1 # truncate to maximum observed value to avoid simulating aberrant values
  return(y)
}
### ZIP Poisson with gender on both intercept and slope, correlation between intercept and slope
covmat<-diag(c(1,1,0))
covmat[1,2]<-covmat[2,1]<-1
saemix.model.zip.cov<-saemixModel(model=count.poissonzip,description="count model ZIP",
  modeltype="likelihood", simulate.function = countsimulate.poissonzip,
  psi0=matrix(c(1.5, 0.01, 0.2),ncol=3,byrow=TRUE,dimnames=list(NULL, 
  c("intercept", "slope","p0"))), transform.par=c(0,0,3), 
  covariance.model=covmat, omega.init=diag(c(0.5,0.3,0)),
  covariate.model = matrix(c(1,1,0),ncol=3, byrow=TRUE), verbose=FALSE)
saemix.options<-list(seed=632545,save=FALSE,save.graphs=FALSE, 
  displayProgress=FALSE, fim=FALSE, print=FALSE)
zippoisson.fit.cov<-saemix(saemix.model.zip.cov,saemix.data,saemix.options)

nsim<-100
yfit1<-simulateDiscreteSaemix(poisson.fit.cov, nsim=nsim)
plot1 <- discreteVPC(yfit1, outcome="count", which.cov="gender", breaks=c(0:9, 16, 25,80))
\end{lstlisting}

\subsection*{Additional results for the RAPI dataset}

%\Eco{Confused with the hurdle model: very different estimates for the logistic regression model compared to Atkins (see notebook compared to Table 2), with an intercept of 2.48 and a time effect of -0.05 (versus 1.56 and -0.03 in the paper, and also different estimates for the gender effects. However our model seems to fit just fine when looking at VPC. Another puzzling difference is that with the hurdle model, using the parameters reported in Atkins for the logistic component seems to yield a higher than expected proportion of 0's at time 0 in women (0.24) compared to ours (0.12) and to the observed data (0.10) (note: no variance reported in Atkins so used our estimate of variance, but even with a lower variance still higher). {\bfseries Question: is the binary part of the hurdle model implemented correctly here ?? but the binary simulation study shows saemix can recover the right estimates} }
% maybe mistake in the original publication...

\begin{figure}[!h]
    \centering
    \includegraphics[scale=0.4]{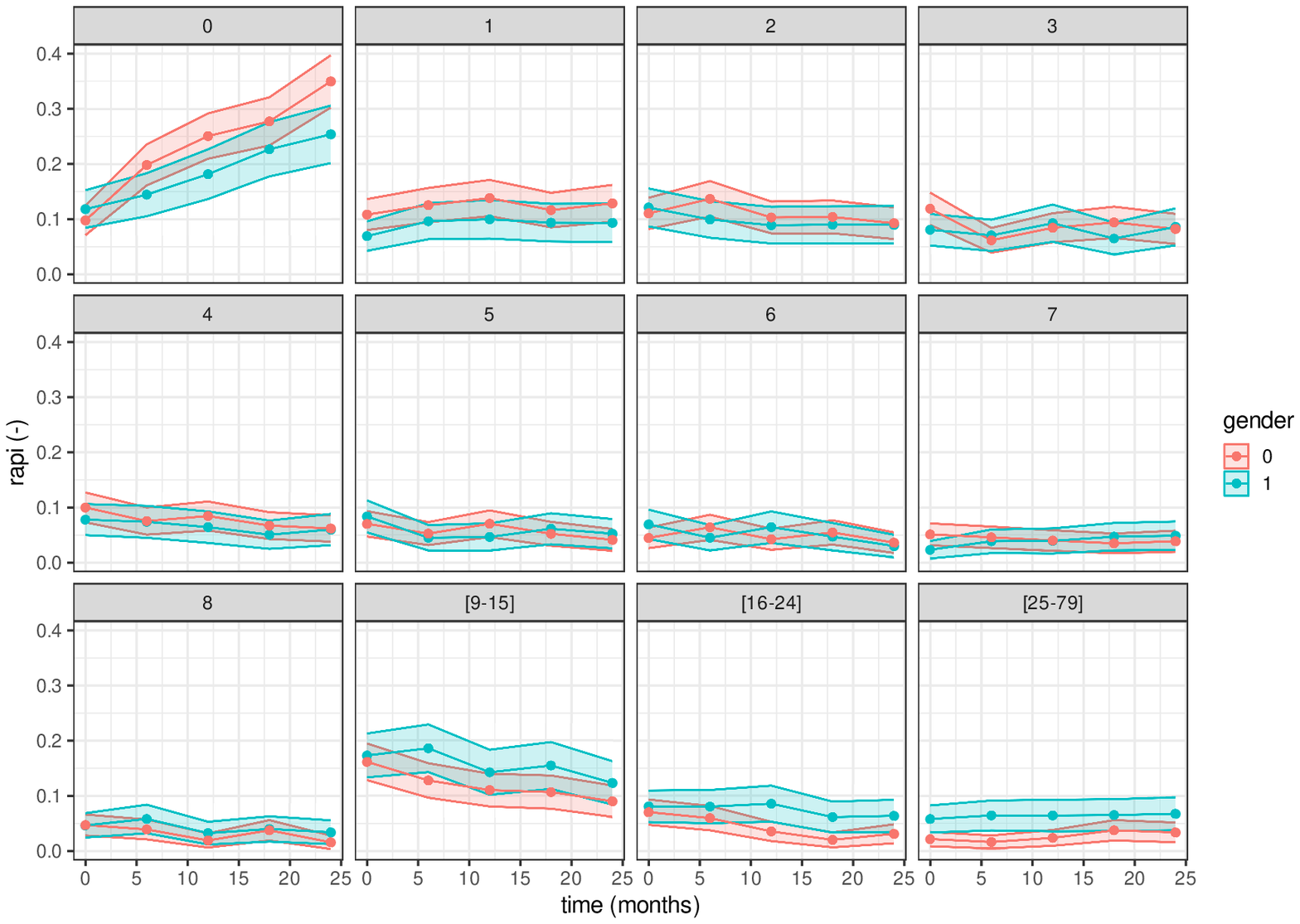}
    \caption{\label{fig:dataRapi} Evolution of the observed proportion of each count over time, stratified by gender. Large counts were regrouped (see facet title).}
\end{figure}

Figure~\ref{fig:dataRapi} shows the proportion of subjects presenting each score as a function of time, stratified by gender. We observe an increase in the proportion of subjects reporting no alcohol-related problem as time goes on, resulting from a slow but steady decrease in the number of binge-episodes reported. We also not a clear gender gap both in the trend and in the number of reported episodes.

Table~\ref{tab:rapiModelsSE} gives the parameter estimates for the three models along with their 95\% confidence intervals estimated using conditional bootstrap. For the hurdle model we show the parameter estimates for the two submodels, logistic regression for zero counts and truncated Poisson for non-zero counts.
\begin{table}[!h]
\begin{center}    
    \begin{tabular}{ccccc}
\hline
  & Poisson & ZIP & \multicolumn{2}{c}{Hurdle} \\ 
  & & & Logistic regression & Truncated Poisson\\
  \hline
$\alpha_0$ (-) &  1.49 (0.05) &  1.57 (0.045) &  2.49 (0.165) &  1.61 (0.05) \\ 
  $\beta_{male,\alpha_0}$ (-) &  0.20 (0.07) &  0.19 (0.07) &  0.01 (0.19) &  0.21 (0.07) \\ 
  $\alpha_1$ (mo$^{(-1)}$) & -0.039 (0.0035) & -0.035 (0.0035) & -0.054 (0.011) & -0.018 (0.003) \\ 
  $\beta_{male,\alpha_1}$ (-) &  0.017 (0.005) &  0.016 (0.005) &  0.038 (0.015) &  0.01 (0.004) \\ 
  $p_0$ (-) & - &  0.076 (0.005) & - & - \\ 
  $\omega_{\alpha_0}$ (-) &  0.92 (0.06) &  0.80 (0.05) &  1.54 (0.47) &  0.72 (0.05) \\ 
  $\omega_{\alpha_1}$ (-) &  0.004 (0.0003) &  0.003 (0.0002) &  0.007 (0.002) &  0.002 (0.0002) \\ 
  $\rho(\alpha_0,\alpha_1)$ (-) & -0.14 (0.047) & -0.085 (0.05) &  0.38 (0.20) & -0.29 (0.055) \\ 
   \hline \\
    \end{tabular}
    \caption{Parameter estimates and their SE estimated by conditional bootstrap for the Poisson, Zero-Inflated Poisson and Hurdle models. \label{tab:rapiModelsSE}}
\end{center}
\end{table} 
% adjusted nb of digits

\begin{figure}[!h]
    \centering
    \includegraphics[scale=0.4]{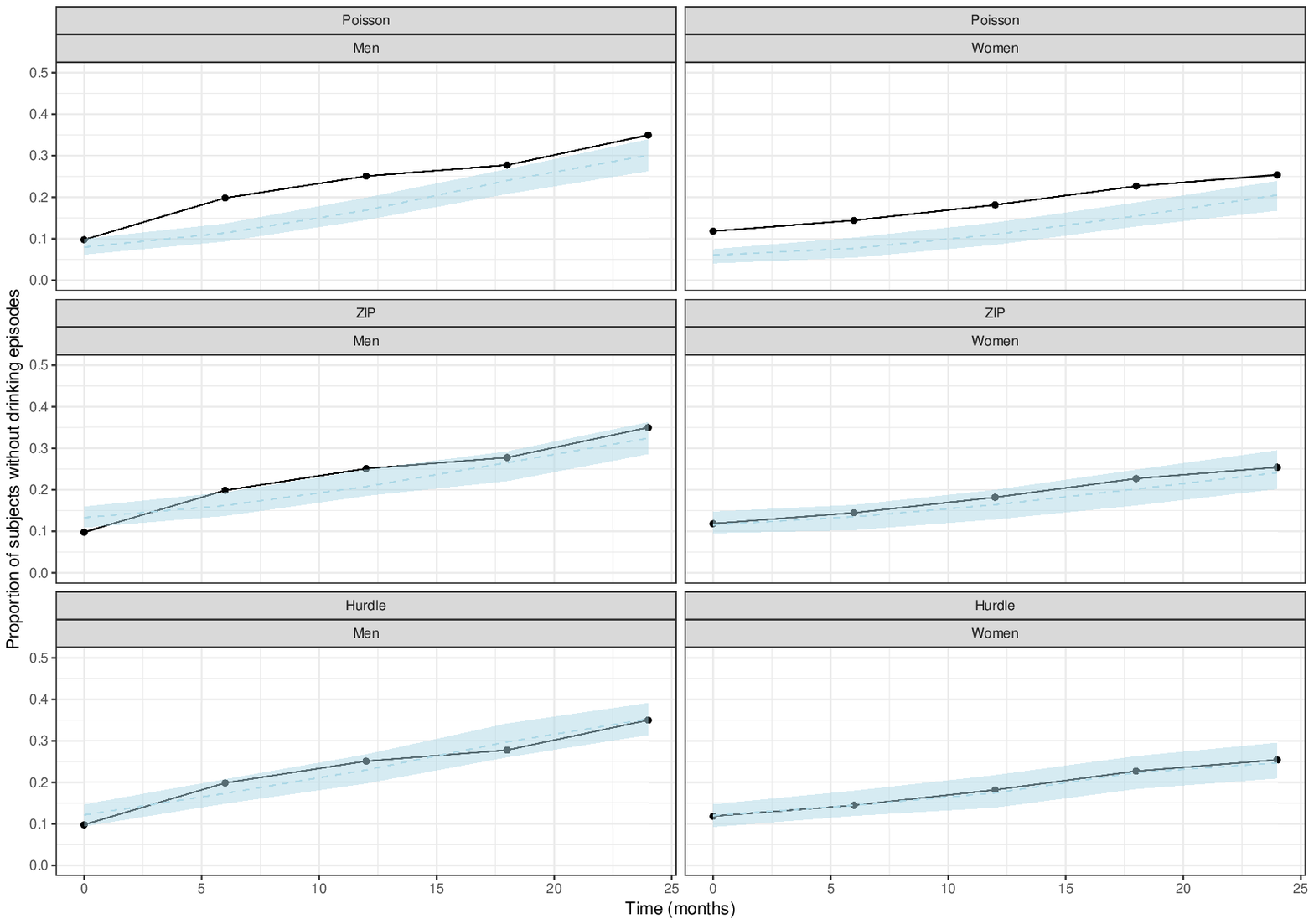}
    \caption{\label{fig:rapiPropZeroes} VPC for the proportion of subjects who didn't report any drinking episode, stratified by gender, for the Poisson, Zero-Inflated Poisson and Hurdle models.}
\end{figure}

Figure~\ref{fig:rapiPropZeroes} compares the evolution of the proportion of zero counts across time, stratified by gender. The VPC for the Poisson model showed an underprediction throughout the study of the proportion of subjects reporting no drinking episode, especially marked in women, while the two other models performed adequately.

\begin{figure}[!h]
    \centering
    \includegraphics[scale=0.6]{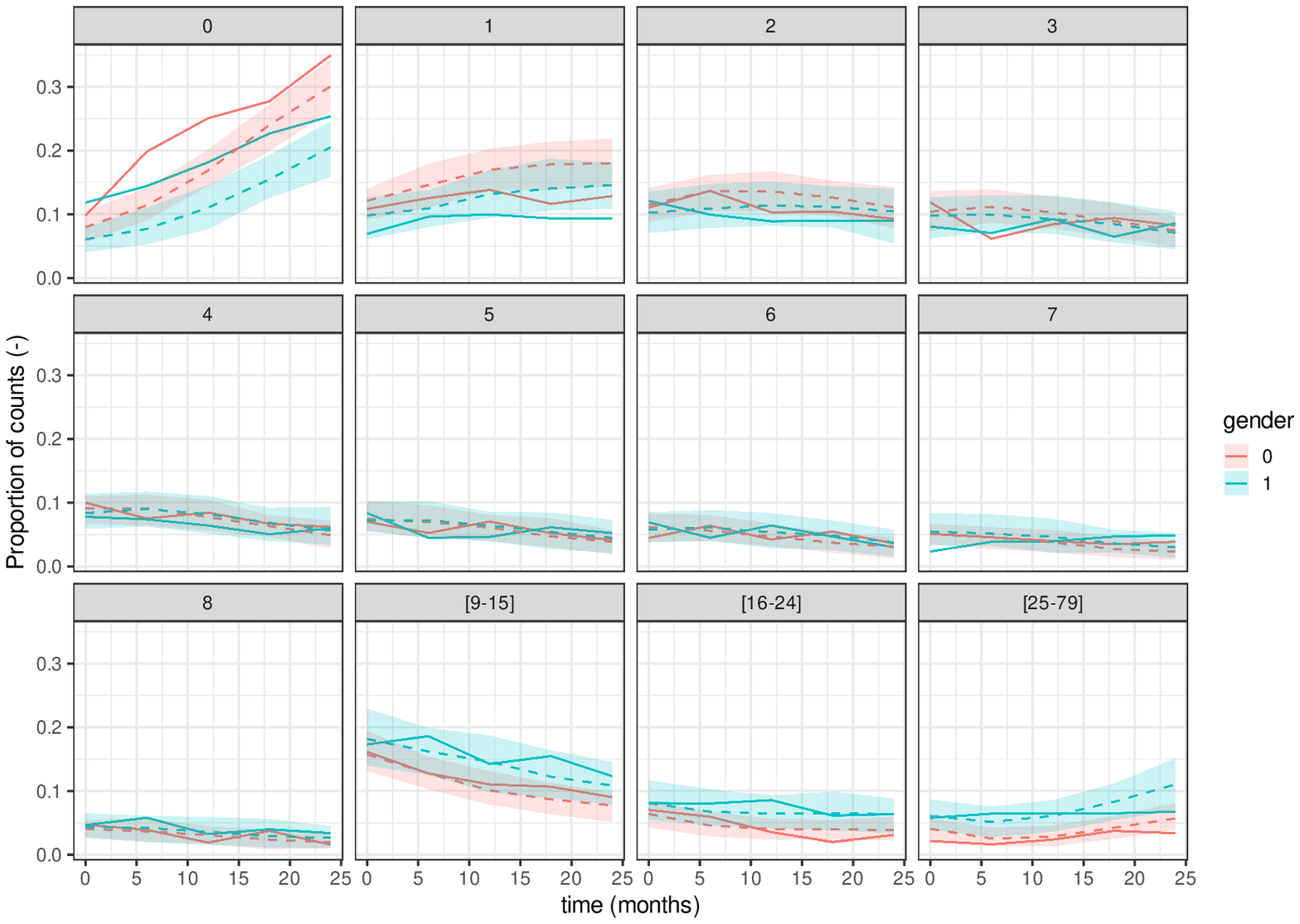}
    \caption{\label{fig:rapiPoissonVPC} VPC plots for the Poisson model, stratified by gender.}
\end{figure}

\begin{figure}[!h]
    \centering
    \includegraphics[scale=0.6]{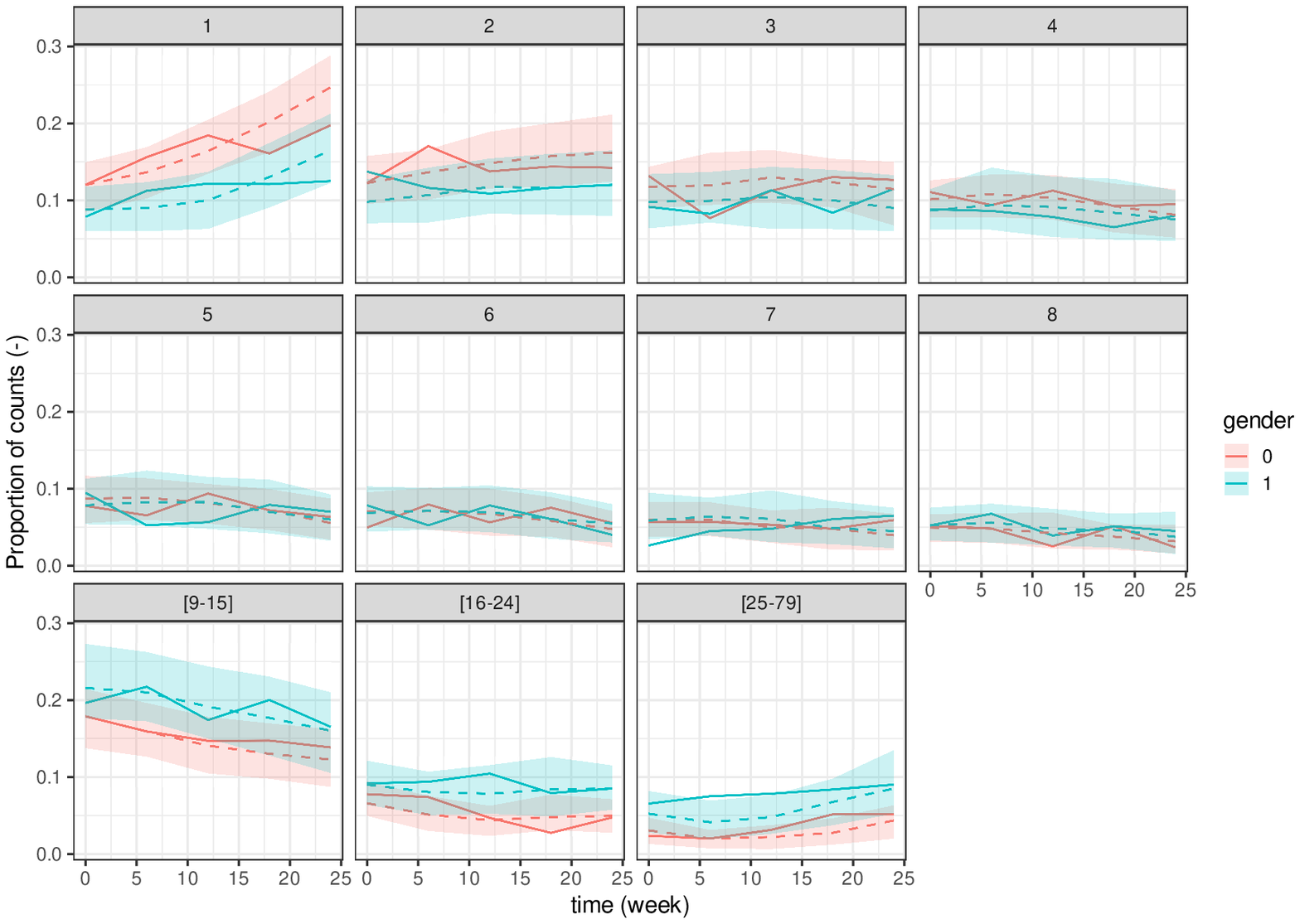}
    \caption{\label{fig:hurdleVPC} VPC plots for the non-zero categories using the Hurdle model, stratified by gender.}
\end{figure}

Diagnostic graphs for the Poisson model are given in Figure~\ref{fig:rapiPoissonVPC}, while the corresponding graphs for non-zero counts of the Hurdle model (Figure~\ref{fig:hurdleVPC}) demonstrate better adequacy across count categories.

\clearpage
\newpage
\section*{Appendix E - Time-to-event models}

\subsection*{saemix code}

\begin{lstlisting}[language=R]
data(lung.saemix)
lung2<-lung.saemix
lung2$ecog1<-ifelse(lung2$ph.ecog==1,1,0)
lung2$ecog23<-ifelse(lung2$ph.ecog>1,1,0)
lung2$pat.karno[is.na(lung2$pat.karno)]<-median(lung2$pat.karno, na.rm=TRUE)
lung.data<-saemixData(name.data=lung2,header=TRUE,name.group=c("id"),
  name.predictors=c("time","status","cens"),name.response=c("status"),
  name.covariates=c("age", "sex", "ecog1","ecog23", "ph.karno", "pat.karno","ph.ecog"),
  units=list(x="days",y="",covariates=c("yr","","-","-","%","%")), verbose=FALSE)
plotDiscreteData(lung.data, outcome="tte", which.cov="sex")

weibulltte.model<-function(psi,id,xidep) {
  T<-xidep[,1]
  y<-xidep[,2] # events (1=event, 0=no event)
  cens<-which(xidep[,3]==1) # censoring times (subject specific)
  init <- which(T==0)
  Te <- psi[id,1] # Parameters of the Weibull model
  gamma <- psi[id,2]
  Nj <- length(T)
  
  ind <- setdiff(1:Nj, append(init,cens)) # indices of events
  hazard <- (gamma/Te)*(T/Te)^(gamma-1) # h
  H <- (T/Te)^gamma # H= -ln(S)
  logpdf <- rep(0,Nj) # ln(l(T=0))=0
  logpdf[cens] <- -H[cens] + H[cens-1] # ln(l(T=censoring time))=ln(S)=-H
  logpdf[ind] <- -H[ind] + H[ind-1] + log(hazard[ind]) # ln(l(T=event time))=ln(S)+ln(h)
  return(logpdf)
}
# Simulate events based on the observed individual censoring time
simulateWeibullTTE <- function(psi,id,xidep) {
  T<-xidep[,1]
  y<-xidep[,2] # events (1=event, 0=no event)
  delta <- xidep[,3] # censoring indicator
  cens<-which(xidep[,3]==1) # censoring times (subject specific)
  tmax <- max(T[cens]) # maximum censoring time observed in dataset
  init <- which(T==0)
  Te <- psi[,1] # Parameters of the Weibull model
  gamma <- psi[,2]
  Nj <- length(T)
  ind <- setdiff(1:Nj, append(init,cens)) # indices of events
  tevent<-T
  Vj<-runif(dim(psi)[1])
  tsim<-Te*(-log(Vj))^(1/gamma) #   events
  tevent[T>0]<-tsim
  tevent[delta==1 & tevent>T] <- T[delta==1 & tevent>T] # subject-specific censoring time
  return(tevent)
}
covmodel <- cbind(c(0,1,0,1,0,0),rep(0,6))
weibull.model.cov2<-saemixModel(model=weibulltte.model,description="Weibull TTE model",
  modeltype="likelihood",psi0=matrix(c(300,2),ncol=2,byrow=TRUE,
  dimnames=list(NULL,  c("Te","gamma"))), transform.par=c(1,1),
  covariance.model=matrix(c(0,0,0,1),ncol=2, byrow=TRUE), 
  covariate.model=covmodel, verbose=FALSE)
saemix.options<-list(seed=632545,save=FALSE,save.graphs=FALSE, 
  displayProgress=FALSE, print=FALSE)
weibcov.fit2<-saemix(weibull.model.cov2,lung.data,saemix.options)

saemix.model<-saemixModel(model=weibulltte.model,description="Weibull TTE model",
  modeltype="likelihood", psi0=matrix(c(1,2),ncol=2,byrow=TRUE,
  dimnames=list(NULL,  c("Te","gamma"))), transform.par=c(1,1),
  covariance.model=matrix(c(1,0,0,0),ncol=2, byrow=TRUE), verbose=FALSE)
saemix.options<-list(seed=632545,save=FALSE,save.graphs=FALSE, 
  displayProgress=FALSE, print=FALSE)
tte.fit<-saemix(saemix.model,lung.data,saemix.options)
plot(tte.fit, plot.type="convergence")
\end{lstlisting}

\end{document}